\documentclass[12pt]{article}

\usepackage[margin=1in]{geometry}
\usepackage{amsmath,amssymb,amsthm}
\usepackage{mathtools}
\usepackage{bm}
\usepackage{graphicx}
\usepackage{xcolor}
\usepackage[colorlinks=true, linkcolor=blue, citecolor=blue, urlcolor=blue, hypertexnames=false]{hyperref}
\usepackage[numbers]{natbib}
\usepackage{setspace}
\usepackage{microtype}
\usepackage{booktabs}
\usepackage{caption}
\usepackage{subcaption}
\usepackage{float}

\newcommand{\Etot}{E_{\mathrm{tot}}}
\newcommand{\Eembed}{E_{\mathrm{embed}}}
\newcommand{\Epair}{E_{\mathrm{pair}}}
\newcommand{\Keff}{K_{\mathrm{eff}}}
\newcommand{\eps}{\varepsilon}
\newcommand{\epszero}{\varepsilon_{0}}
\newcommand{\xvec}{\mathbf{x}}
\newcommand{\cvec}{\mathbf{c}}

\newcommand{\calJ}{\mathcal{J}}
\newcommand{\calN}{\mathcal{N}}
\newcommand{\calO}{\mathcal{O}}
\newcommand{\Ebar}[1]{\overline{\widetilde{E}}_{#1}}
\newcommand{\Etilde}[1]{\widetilde{E}_{#1}}
\newcommand{\Abar}[1]{\overline{A}_{#1}}
\newcommand{\Shat}{\widehat{S}}
\newcommand{\Hhat}{\widehat{H}}
\newcommand{\rmin}{r_{\min}}
\newcommand{\dist}{\mathrm{dist}}

\title{\textbf{Scaling atom-by-atom inverse design with nano-topology
optimization and diffusion models}}

\author{Chun-Teh Chen$^{1}$ \and Denvid Lau$^{2}$}

\date{
  \small $^{1}$Department of Materials Science and Engineering,
  University of California,\\
  Berkeley, CA, USA\\[2pt]
  \small $^{2}$Department of Architecture and Civil Engineering,
  City University of Hong Kong,\\
  Hong Kong, China\\[4pt]
  \small Correspondence: \texttt{chunteh@berkeley.edu}}

\begin{document}

\maketitle

\begin{abstract}
The mechanical properties of metallic nanostructures are governed not
only by topology but also by crystal symmetry and face-specific surface
physics, which are typically absent from continuum topology optimization. We
develop an atom-by-atom inverse design framework that combines
Nano-Topology Optimization (Nano-TO) with conditional denoising
diffusion probabilistic models. Nano-TO treats each atom as a discrete
design variable and evaluates stiffness from the symmetric curvature of
the total energy, removing residual surface-stress bias. A
crystallography-aligned multi-shell sensitivity filter stabilizes the
optimization and enables designs containing more than $6.5\times10^{5}$
atoms. Using aluminum nanocantilevers, we identify a
surface-physics-driven topology selection rule: thickness-periodic beams
favor brace-dominated trusses, whereas finite-thickness beams favor
nearly closed walls that provide efficient shear paths and reduce surface
penalties. At sufficiently small scales, these walls become mechanically
unstable, and truss-like layouts reappear. In nanopillar studies, atomistic
optimization outperforms continuum topology-optimized designs. Finally,
conditional diffusion models trained on Nano-TO data generate diverse
high-performance candidates near the optimization frontier. These results
establish nanoscale inverse design as a coupled problem of topology and
surface physics.
\end{abstract}

\newpage

At nanometer scales, topology not only influences how forces are
transmitted through a nanostructure but also dictates which
crystallographic facets, edges, and low-coordination atomic sites are
exposed. This is especially important in micro- and
nano-electromechanical systems (MEMS/NEMS), including resonators,
sensors, and scanning probes, whose performance depends on elastic
response~\cite{Rugar2004,Ekinci2005}. At macroscopic scales, continuum elasticity
is usually adequate as atomic details can be averaged without
significantly altering the predicted behavior. At nanoscale dimensions,
however, a large fraction of atoms are located at free surfaces, where
reduced coordination, relaxation, and facet-dependent bonding alter
residual stress and elasticity. Mechanical response, therefore, depends
jointly on topology, crystal symmetry, and face-specific surface
physics~\cite{Trimble2003,Deng2009,Shenoy2005}. Experiments and atomistic
simulations on nanowires have shown pronounced size effects in
the effective axial modulus, with the modulus either decreasing or
increasing with radius, depending on material, wire orientation, and
exposed facets~\cite{Zhang2008,Wang2008,Zhu2012,Miller2000,Cuenot2004}. Nanoscale inverse
design must therefore optimize not only the distribution of atoms but
also the atomic surfaces this topology creates.

Topology optimization (TO) provides a powerful framework for structural
layout design~\cite{Bendsoe1988,Bendsoe1989,Bendsoe2013,Eschenauer2001,Sigmund2013},
and density-based implementations now scale to very large continuum
problems~\cite{Aage2017,Andreassen2011,Liu2014}. Standard TO, however,
treats the solid as a homogeneous medium. It optimizes the coarse
geometry of a structure, but it does not specify which crystallographic
facets, edges, or local atomic motifs are created by that geometry.
Continuum extensions based on surface elasticity, including
Gurtin--Murdoch models~\cite{Gurtin1975,Zhu2017,Nanthakumar2015}, and
higher-order theories such as strain-gradient and couple-stress
formulations~\cite{Lam2003,Mindlin1965}, can capture partial size
effects by introducing effective surface constitutive laws and intrinsic
length scales. These approaches have been valuable for predicting how
surfaces change the mechanical responses of prescribed nanostructures
and, in some cases, for incorporating surface effects into
continuum-level shape or topology optimization. However, they still
describe both bulk and surface in homogenized form. Consequently, these
approaches cannot directly resolve atomistic realizations of a
coarse-grained surface orientation, such as surface terminations, atomic
steps and terraces, local coordination changes, or discrete defects.
This limitation becomes especially important in atomistic inverse design,
as changing the topology at the nanoscale simultaneously alters the
populations of exposed facets, edges, and low-coordination sites that
jointly determine target mechanical responses.

Nano-Topology Optimization (Nano-TO) addresses this atomistic
inverse-design gap by treating each atom as a discrete design
variable~\cite{Chen2020NanoTO}. Rather than predicting surface effects
for a fixed geometry, Nano-TO allows topology and the surfaces created
by that topology to be determined together. This formulation can, in
principle, resolve the discrete surface and lattice physics that
continuum models omit. In practice, however, atomistic inverse design is
much harder to scale. Each design update must be evaluated through
nonlinear relaxations under interatomic potentials, and the resulting
per-atom sensitivities become noisy and unstable if used too locally.
This instability limits accessible system size and can disrupt the
formation of coherent load-bearing paths. A second challenge is
non-uniqueness. For a given set of target properties, there is generally
not a single admissible nanostructure, but rather a family of distinct
atomistic topologies with comparable performance. Deterministic
optimization can find one feasible design, but it does not, by itself,
map the broader near-optimal design manifold or expose useful trade-offs
among the properties of interest.

We address both limitations by combining atomistic topology optimization
with generative modeling. Building on our earlier Nano-TO framework~\cite{Chen2020NanoTO}, we formulate stiffness through
the symmetric energy-curvature measure that removes residual
surface-stress bias, and we introduce a crystallography-aligned
multi-shell sensitivity filter that regularizes per-atom sensitivities
sufficiently to enable stable large-batch Nano-TO in systems exceeding
$6.5\times10^{5}$ atoms. More broadly, filtering and minimum-length-scale
control have long served as regularization tools in
TO~\cite{Sigmund2007,Guest2004,Lazarov2011}, although here the neighborhood is chosen
to reflect crystallography connectivity and the interaction range of the
interatomic potential. We then couple the resulting optimization data to
conditional denoising diffusion probabilistic models
(c-DDPMs)~\cite{SohlDickstein2015,Ho2020,Song2021}. Recent studies have applied
generative models, including diffusion-based models, to inverse design
problems~\cite{Chen2019backprop,Kang2024,SanchezLengeling2018,Zheng2023truss,Mao2020GAN,Bastek2023,Li2026}.
We use c-DDPMs to sample a diverse set of target-consistent, near-optimal
designs. Using aluminum nanocantilevers and nanopillars as testbeds, we
show that explicit surface physics can qualitatively change the optimal
topology: thickness-periodic cantilevers favor truss-like motifs, exposed
side surfaces drive nearly closed-wall designs, and at a smaller scale,
the optimum shifts back toward truss-like layouts as ultrathin walls lose
their ability to carry transverse shear at the nanoscale. These
results establish a route to inverse design in which topology and surface
physics are optimized simultaneously, while generative models broaden
access to high-performing alternatives and multi-objective design
trade-offs.

\section*{Results}

\subsection*{Nano-TO and c-DDPM frameworks}

To make surface physics part of the inverse-design problem, we represent
each nanostructure atom by atom and evaluate its mechanical properties with an
embedded-atom method (EAM) potential~\cite{Mishin1999,Daw1984}. This description
captures the facet-dependent surface elasticity of FCC metals, whose
low-coordination surfaces exhibit distinct symmetries and in-plane elastic
responses (Supplementary Notes A.1, Supplementary Figures S1 and S2).
Changes in topology modify not only the load path, but also the
populations of surfaces that contribute to stiffness. Figure~1a
summarizes the resulting Nano-TO workflow, which builds on our previous
atom-by-atom inverse materials design formulation~\cite{Chen2020NanoTO}.
For a prescribed loading mode, stiffness is evaluated from the symmetric
curvature of the total energy, which removes the linear contribution from
residual surface stress (Methods). Atomistic relaxation makes the per-atom sensitivities noisy and highly
local, especially as system size and geometric complexity increase. We
address this with a crystallography-aligned multi-shell sensitivity
filter spanning the first 12 FCC shells (Supplementary Notes A.2,
Supplementary Table S1, Supplementary Figure S3). This multi-shell filter
suppresses atom-scale fluctuations while preserving coherent load-bearing
paths, enabling stable, large-batch updates in which atoms are removed
from low-contributing sites and restored at favorable virtual sites. The
resulting stabilization makes atomistic inverse design practical for
systems with more than $6.5\times10^{5}$ atoms. By contrast, a
first-shell local filter yields unstable optimizations, with disconnected
void networks and failure to reach the target property (Supplementary
Notes A.3, Supplementary Figure S4).

We then use c-DDPMs as a complementary, data-driven layer that learns a
property-conditioned distribution over nanostructures. Figure~1b
summarizes the proposed c-DDPM workflow. For the beam problems
considered, each design is encoded as a binary cross-sectional image and
labeled by quantities evaluated from atomistic simulations (e.g., mass
ratio, effective stiffness). These conditioning variables are embedded
and fed into the network through cross-attention layers. At each
attention block, the conditioning embedding serves as a set of keys and values
that the spatial feature queries attend to, enabling the denoiser to
modulate its reconstruction based on the target property (Methods).
During training, the network learns to progressively remove noise from
perturbed images by minimizing a reconstruction loss under this
conditioning. At inference, the model starts from random noise and, with
classifier-free guidance (CFG)~\cite{Dhariwal2021,Ho2022CFG}, generates multiple
candidates consistent with the specified targets rather than a single
deterministic solution. As used below, this framework serves two
purposes: it provides a gradient-free inverse-design benchmark when
trained on broad synthetic samples, and it explores diverse near-optimal
candidates when trained on Nano-TO output.

\subsection*{Design of nanocantilevers under thickness-periodic boundary
conditions using Nano-TO}

We first examine aluminum nanocantilevers under thickness-periodic
boundary conditions, which suppress side surfaces and provide a
controlled benchmark aligned with the two-dimensional cross-sectional
representation used later for c-DDPMs. Aluminum is a useful model system
as its bulk elasticity is nearly isotropic, making deviations from
simple continuum scaling easier to attribute to topology and surface
effects. The design domain measures 200.475$\times$20.25$\times$615.60~\AA\
and contains 150{,}480 atoms, including 148{,}500 active atoms and
1{,}980 passive atoms at the clamped boundary. A vertical displacement is
applied at the mid-plane of the free end, and the objective is to
minimize bending compliance at prescribed mass ratios. Since the
geometry, loading, and material are invariant under reflection about the
mid-plane, we enforce mirror symmetry to reduce the design space
(Supplementary Notes A.4, Supplementary Figure S5). For each target mass
ratio, 64 independent trials are performed from a fully dense beam (mass
ratio of 100\%), with different optimization paths initiated by a small
random perturbation before each energy minimization. Optimization then
proceeds through a mass-reduction stage followed by mass-conserving
refinements (Methods).

The optimized designs and performance are shown in Figure~2. Nano-TO
does not simply reduce the beam height uniformly. Instead, it
consistently generates truss-like designs with multiple cross-braces
(Figure~2a), with internal voids opening while a connected network of
inclined members is retained between the clamp and the loaded end. We
quantify performance by the bending stiffness normalized by that of the
fully dense beam and report all 64 trials at each mass ratio
(Figure~2b). Across the full mass-ratio range, the optimized designs
outperform the uniformly height-scaled reference beams of equal mass
(Supplementary Notes A.5, Supplementary Figure S6). The stiffness of
these reference beams closely follows the Euler--Bernoulli estimate,
consistent with the translationally periodic geometry and relatively low
surface-to-volume ratio of this benchmark. At a mass ratio of 59.60\%,
the best Nano-TO design retains a normalized stiffness of 0.820, whereas
the corresponding reference beam reaches only 0.223. Thus, removing more
than 40\% of the atoms reduces stiffness by only 18\% in the optimized
design, compared with more than 77\% in the reference beam. This
thickness-periodic case establishes the baseline topology preferred when
side-surface atoms are absent.

The preference for cross-braced layouts is robust to symmetry
constraints. When mirror symmetry is removed, Nano-TO consistently
generates related truss-like designs, and the best design at a mass
ratio of 59.60\% reaches a normalized stiffness of 0.818 from 64 trials,
only 0.26\% below the mirror-symmetric case (Supplementary Notes A.6,
Supplementary Figures S7 and S8). The small difference suggests that the
symmetry constraint mainly improves search efficiency rather than
changing the accessible optimum. During optimization, we occasionally
observe transient drops in stiffness (Figure~2c), which coincide with
pattern transitions that temporarily create disconnected floating atoms.
These atoms contribute to mass but not to load transfer. Subsequent
iterations identify these atoms and remove them from the design space,
restoring the expected performance trend.

\subsection*{Design of nanocantilevers under thickness-periodic boundary
conditions using c-DDPM}

Having established the Nano-TO performance frontier for
thickness-periodic nanocantilevers, we next examine whether conditional
diffusion models can recover high-stiffness designs under the same
setting. We fix the mass ratio at 59.60\% and use c-DDPMs in two
complementary ways: first as a purely data-driven inverse-design
benchmark trained on generic synthetic layouts, and then as a sampler of
diverse near-optimal candidates trained on Nano-TO outputs. In both
cases, generated designs are converted to atomistic models and evaluated
using the same bending-stiffness calculation as in the preceding section
(Methods).

The first model, Gaussian-DDPM, is trained on 32{,}000 valid layouts
generated from Gaussian random fields (GRFs), which provide a broad
geometric prior but are not mechanics-informed~\cite{Bastek2023,Li2026}. The
training set spans a wide performance range, with a mean normalized
stiffness of 0.116 and a maximum of 0.604. Under a high-stiffness
conditioning label and classifier-free guidance (Methods, Supplementary
Notes A.7, Supplementary Figure S9), denoising gradually produces smooth,
curved motifs characteristic of the GRF prior (Figure~3a, left). The
generated samples exhibit a substantial shift toward higher performance
relative to the training set (Figure~3a, right). The mean normalized
stiffness is 0.616, which is 5.3 times that of the training samples, and
the best generated design reaches 0.736, more than 21\% above the best
training sample. Notably, the mean stiffness of the generated designs
even exceeds the maximum stiffness in the training samples, indicating
that conditioning and guidance enable extrapolative sampling rather than
memorization of the training samples. Nevertheless, the best
Gaussian-DDPM design remains below the best Nano-TO design at the same
mass ratio (0.736 versus 0.820), indicating that a generic smooth-layout
prior does not fully recover the brace-dominated load paths favored by
atomistic optimization.

The second dataset is compiled from Nano-TO outputs. These samples are
not specifically created for training a c-DDPM but are instead reused
from previous optimization runs. At a mass ratio of 59.60\%, only 64
Nano-TO designs are available, which are too few to train a c-DDPM. To
address this data shortage, we train a model on various mass ratios and
condition it on the desired value at inference. The idea is that exposing
the model to designs with different mass ratios helps it understand how
geometry determines load paths and affects bending stiffness. From an
original pool of 48{,}000 Nano-TO designs, we retain 32{,}000 after
removing disconnected layouts. Each sample is labeled with its mass
ratio and bending stiffness. We refer to the model trained on this
dataset as TO-DDPM. The inference condition is chosen based on a grid
search over stiffness conditioning and guidance strength at the target
mass ratio of 59.60\% (Methods, Supplementary Notes A.8, Supplementary
Figure S10). In contrast to Gaussian-DDPM, denoising rapidly organizes
the layouts into truss-like motifs that closely resemble the Nano-TO
designs (Figure~3b, left), showing that the learned prior is already
aligned with the underlying mechanics. In a production run of 32{,}000
samples under the chosen inference setting, the generated samples lie
within a narrow high-performance band (Figure~3b, right), with a mean
normalized stiffness of 0.809, and the best design reaches 0.860. Within
the current optimization budget, diffusion models complement Nano-TO by
efficiently exploring the learned near-optimal design space.

To understand how TO-DDPM constructs these candidates, we compare
generated designs with the Nano-TO training set using percent identity
over the active atoms (Supplementary Notes A.9). The best design DM-22397
has nearest-neighbor similarities of 93.55\% and 89.39\% to TO-46 and
TO-06, respectively (Figure~4a). The generated beam is a composite:
roughly the first third reproduces the topology of TO-06, while the
remaining two-thirds follow TO-46, with both regions showing fewer
atomic modifications in their respective overlays. DM-24015, by
contrast, has similarities of 89.76\% and 89.54\% to TO-57 and TO-23
(Figure~4b). However, only a localized region resembles its nearest training
neighbor. Together, the two cases illustrate that TO-DDPM operates along
a spectrum: from recombining recognizable sub-structures to
synthesizing globally new topologies informed by the full training
distribution. The resulting local edits increase stiffness by 1.23\%
to 6.00\% relative to the nearest training samples.
Across the generated set, no two designs are the same at the atomic
level, which is helpful for downstream screening. As one example,
TO-DDPM produces a design with a normalized stiffness of 0.822, 0.28\%
above the best Nano-TO design, while reducing the surface-atom fraction
from 0.1436 to 0.1367 and lowering the potential energy per atom by
0.0012~eV (Supplementary Notes A.10, Supplementary Figure S11). Diffusion
models, therefore, do not replace Nano-TO; rather, they expand a single
optimized solution into a family of high-performing alternatives that can
be screened under additional criteria.

\subsection*{Design of finite-thickness nanocantilevers using Nano-TO}

We next solve the finite-thickness nanocantilever problem with Nano-TO
by removing the thickness-periodic boundary condition and tripling the
beam thickness, while keeping the loading and optimization protocol
unchanged. In both the thickness-periodic and finite-thickness cases,
the mechanical response is evaluated using three-dimensional atomistic
models. Under thickness periodicity, the topology is constrained to
remain extruded through the thickness, whereas in the finite-thickness
problem, the side surfaces are exposed, and atoms can be redistributed
along the thickness direction (Supplementary Notes A.11). Starting again from a fully dense beam,
Nano-TO converges to nearly closed-wall designs (Supplementary Figure S12). Across the mass ratios studied, the optimized designs remain
substantially stiffer than height-scaled reference beams of equal mass
(Supplementary Figure S13). The optimized design, with a mass
ratio of 59.60\%, is shown in Figure~5a.

To isolate the contribution of the newly exposed side surfaces, we
construct a controlled baseline by taking the thickness-periodic design
from Figure~2a, tripling its thickness, and removing the periodic
boundary conditions, while keeping its in-plane topology unchanged. This
side-exposed truss-like structure is no longer optimal; rather, it
reflects the penalty incurred when the extruded truss is exposed to free
side surfaces. Its effective bending stiffness decreases by approximately
13\%, while the surface-atom fraction increases from 13.79\% to 19.54\%
(Supplementary Notes A.12, Supplementary Table S2). This loss reflects a
side-surface penalty at this scale. The exposed side area of this
baseline is dominated by coordination-8 atoms, associated primarily with
\{100\}-like surfaces (Supplementary Figure S14a, b).

Optimizing the finite-thickness problem leads to a different solution.
At the same mass ratio of 59.60\%, the optimized nanocantilever lowers
the surface-atom fraction to 15.51\% and is 5.29\% stiffer than the
side-exposed truss baseline. Locally, Nano-TO preferentially
exposes \{111\} facets, the stiffest 
FCC surfaces, in place of the \{100\}-dominated surfaces found on the baseline (Figure~5a). Atomistic
strain maps further show that the nearly
closed wall distributes transverse shear through a continuous shell,
whereas the side-exposed truss baseline concentrates shear near brace
junctions and window tips (Supplementary Figure S14c--f). The closed-wall
morphology retains the continuum advantage of
providing an efficient transverse-shear path in bending-dominated
structures~\cite{Rieser2023,Sigmund2016}. At the nanoscale, however, the
same morphology gains an additional benefit absent from continuum
descriptions: it also reduces the surface penalty that weakens the
side-exposed truss baseline. The closed-wall design is therefore
selected by both coupled load-transfer and surface exposure effects.

This preference is, nevertheless, size dependent. When all beam
dimensions are scaled down to about 40\% of the original beams, Nano-TO
still favors nearly closed-wall designs at higher mass ratios. However,
at lower mass ratios, the optimized design reverts toward a truss-like
structure (Supplementary Notes A.13, Supplementary Figures S15 and S16).
The optimized design, with a mass ratio of 60.11\%, is shown in
Figure~5b. At this smaller scale, the wall is reduced to only a few
atomic layers and no longer behaves as a mechanically stable load-bearing
shell. Unlike in continuum TO, where an extremely thin wall remains an
admissible feature, the atomistic wall becomes an unstable carrier of
transverse shear, and Nano-TO redirects load through inclined
cross-braces. The finite-thickness problem is thus governed by a
competition between the benefit of reducing unfavorable surface exposure
and the atomic-scale stability limit of a continuous wall.

\subsection*{Design of nanopillars using Nano-TO}

We lastly test Nano-TO on a nanopillar, providing a loading geometry
distinct from the cantilever problems above and representative of
nanoscale mechanical testing. The design domain measures
162.00$\times$162.00$\times$413.10~\AA\ and contains 652{,}800 atoms,
of which 640{,}000 are active and 12{,}800 are passive atoms from the
clamped base. The pillar is supported at the four corners of the base
and loaded by a vertical displacement applied at the center of the top
surface. Since the target mass ratio is only 20.25\%, Nano-TO is
initialized from a uniform prismatic column at that mass ratio rather
than from a fully dense block, thereby skipping the mass-reduction
stage. Figure~6a shows the optimized designs after 100, 1{,}000, and
3{,}000 iterations. From an initially uniform prismatic column, it
develops four curved corner legs that widen toward the base, forming
``roots'' that enhance load transfer and reduce stress concentrations.

This morphological change produces a large stiffness gain. The vertical
stiffness increases to 3.17 times its initial value after 100 iterations
and to 3.65 times after 3{,}000 iterations, with similar trends across
16 independent trials, as shown in Figure~6b. Notably, this improvement
occurs despite the surface-atom fraction increasing from 11.25\% in the
initial uniform prismatic column to 19.62\% in the optimized design.
Thus, Nano-TO does not maximize stiffness by minimizing total surface
area. Instead, it generates more surface while rearranging both the global load
path and the local surface orientations in a mechanically favorable way.

Figure~6c summarizes this change in surface character using coordination
numbers as proxies for local surface orientations. The initial structure
is dominated by coordination-8 atoms, which are mainly associated with
\{100\}-like surfaces. During optimization, the fractions of
coordination-7 and coordination-9 atoms, corresponding to \{110\}-like
and \{111\}-like surfaces, rise substantially and eventually exceed the
coordination-8 fraction. These coordination numbers are not exact facet
labels, as stepped and higher-index surfaces can combine near \{111\}
terraces with near \{110\} steps. Despite this ambiguity, the trend
suggests that the optimized design is no longer dominated by broad
\{100\}-like facets but instead contains a mixture of stiffer
\{111\}-like terraces with angled \{110\}-like sections to accommodate
both normal and shear stresses. Surface elasticity helps explain this
shift; however, it does not determine the optimum on its own. Although
\{111\} surfaces offer higher stiffness in aluminum (Supplementary Notes
A.1), the pillar cannot be designed by maximizing \{111\} exposure alone.
It must also transmit load efficiently from the top contact into the four
corner supports. The optimized design reflects a balance
between local surface-elastic advantages and the global three-dimensional
load transfer required by the loading and boundary conditions.

A continuum benchmark supports this interpretation. When the same
nanopillar problem is first solved by FEM-based TO and the resulting
layout is mapped onto the same FCC lattice, the design reaches a
normalized vertical stiffness of 3.44, compared with 3.65 from Nano-TO,
a difference of 6.10\% (Supplementary Notes A.14, Supplementary Figures
S17 and S18). Subsequent Nano-TO refinement of the FEM-TO design increases
the stiffness to 3.70, indicating that continuum TO captures a broadly
reasonable macro-shape, but not the atomically resolved surface structure
that determines the best performance after discretization (Supplementary
Figure S19). The nanopillar case extends the framework beyond
beam bending and shows that atomistic inverse design can improve stiffness
by co-optimizing global load transfer and the local surface orientations
created by the topology.

\section*{Discussion}

This work transforms surface physics
from a forward-prediction correction into a design variable. In most
nanoscale elasticity studies, researchers examine how a prescribed beam,
wire, or pillar is stiffened or softened by its exposed
facets~\cite{Shenoy2005,Zhang2008,Wang2008,Zhu2012,Miller2000,Cuenot2004}. Here,
the facet population is allowed to change, making inverse design a
coupled problem of load-path selection and surface creation. The
symmetric energy-curvature objective eliminates residual surface-stress
bias, allowing the optimizer to focus on tangent stiffness rather than
prestress. Additionally, the crystallography-aligned multi-shell filter
regularizes the atomistic update field at a length scale that maintains
mechanically coherent features. Its significance is not about filtering
alone, nor is it a universal claim that first-shell filtering is always
inadequate. Instead, these results show that large-scale, large-batch
Nano-TO demands stronger, physically aware regularization than smaller
proof-of-concept problems in earlier work~\cite{Chen2020NanoTO}.

The nanocantilever studies uncover a nanoscale topology-selection rule
not observed in standard continuum TO. When thickness periodicity is
present, the admissible structures remain extruded through the thickness,
and the main question is how transverse shear is redirected into axial
load; in this case, brace-dominated motifs are often preferred. Removing
thickness periodicity changes the design problem in two coupled ways:
side surfaces become mechanically active, and atoms can be redistributed
throughout the thickness. The nearly closed-wall solution should
therefore not be interpreted as resulting solely from surface exposure.
Instead, it extends continuum arguments for closed sections in
bending-dominated structures~\cite{Rieser2023,Sigmund2016} by showing
that, at the nanoscale, topology is selected jointly by global load
transfer and local surface physics. The reduced-scale finite-thickness
study makes this point sharper. Once the wall is reduced to only a few
atomic layers, it no longer behaves as a mechanically robust shell. The
return to a truss-like motif identifies an atomistic
stability threshold that has no direct continuum analogue, where an
arbitrarily thin wall can remain an admissible feature.

The results of generative models are most useful when interpreted in
terms of a map of the design space, not as a competition for the highest stiffness. The
difference between Gaussian-DDPM and TO-DDPM shows that conditional
generation alone is not sufficient; the learned structural prior is
important. A generic smooth prior learned from GRFs can improve average
performance, but it does not automatically recover the topology class
favored by the mechanics. Training on Nano-TO outputs, however, provides
the model with access to reusable local motifs around brace junctions,
openings, and free-end regions. This allows it to sample nearby
alternatives with similar stiffness but different surface fractions and
energies. This indicates that the relevant object at the nanoscale is
not a single optimum but a narrow, learnable family of nearly optimal
designs. In that sense, Nano-TO identifies high-performing basins while
DDPMs explore their local manifold.

The nanopillar problem extends the same logic beyond beam bending,
highlighting a practical multiscale workflow. The best pillar is not
obtained by minimizing total surface area nor by maximizing any single
favorable facet family. Instead, Nano-TO creates additional surface while
reorganizing both global topology and local surface orientations to
accommodate how load enters through the top contact and is redirected
into the four supports. The comparison with FEM-based TO demonstrates
that continuum TO recovers a reasonable macro-scale load path, but once
that topology is mapped onto an FCC lattice, unresolved surface
realization becomes part of the mechanical error. The improvement
obtained by Nano-TO refinement of the FEM-based TO design suggests that continuum
and atomistic inverse design are better viewed as complementary stages
within a single workflow: continuum TO for efficient macro-topology
generation, followed by atomistic refinement when performance depends on
the surfaces created by that topology.

\section*{Methods}

\subsection*{Atomistic modeling, visualization, and analysis}

All atomistic models are three-dimensional. We study two
boundary-condition settings for nanocantilevers: thickness-periodic, in
which the out-of-plane direction is periodic and the side surfaces are
absent, and finite-thickness, in which the side surfaces are exposed.
Structures are generated from conventional FCC unit cells of aluminum
with lattice parameter $a_0 = 4.05$~\AA, with the cube edges aligned
with the simulation-cell axes. For nanocantilevers, the beam axis is taken as the $z$-direction, the
bending direction as $x$, and the thickness direction as $y$. For
nanopillars, the loading direction is $z$.

All simulations use the Mishin embedded-atom method (EAM) potential for
aluminum~\cite{Mishin1999,Daw1984}, evaluated with Large-scale
Atomic/Molecular Massively Parallel Simulator
(LAMMPS)~\cite{Plimpton1995,Thompson2022}. Atomic positions are relaxed at 0~K under
fixed boundary conditions using the conjugate gradient minimizer, with
energy tolerance $0.0$~eV, force tolerance $10^{-10}$~eV/\AA, a maximum
of 100{,}000 minimization iterations (10{,}000 for nanopillars), and a
maximum of 1{,}000{,}000 force evaluations (100{,}000 for nanopillars).
All three loading states used to evaluate stiffness, $+\epszero$, 0,
and $-\epszero$, are fully relaxed before the corresponding energies are
recorded.

We represent the design space by the binary variable $x_i \in \{0,1\}$,
with $x_i = 1$ for a real atom and $x_i = 0$ for a virtual atom. Within
each design domain, atoms are partitioned into active and passive sets.
Active atoms form the designable region and may switch between real and
virtual states during optimization. Passive atoms remain real throughout
and represent the clamped support or anchored base made of the same
material. In the LAMMPS implementation, virtual atoms provide neither
pair interactions nor electron-density contributions and can later be
reactivated. Passive atoms are held fixed in all Cartesian directions at
the clamped support or anchored base.

Visualization is performed using the Open Visualization Tool
(OVITO)~\cite{Stukowski2009} and Blender. Coordination analysis and
atomic strain analysis are conducted using OVITO.

For the nanocantilever problems, a vertical displacement of magnitude
$\delta_{\mathrm{beam}} = 0.5\%$ of the beam length is applied to the
free end. For the nanopillar problems, a vertical displacement of
magnitude $\delta_{\mathrm{pillar}} = 1\%$ of the pillar height is
applied to the center of the top surface. In the equations below, the
scalar strain amplitude is denoted by $\epszero$ and is related to the
imposed displacement by $\epszero = \frac{\delta}{L_{\mathrm{ref}}}$, where
$L_{\mathrm{ref}}$ is the beam length or pillar height.

The total EAM energy of a structure at an imposed small strain $\eps$ is

\begin{equation}
  \Etot\!\left(\eps;\xvec\right)
  = \Eembed\!\left(\eps;\xvec\right)
  + \Epair\!\left(\eps;\xvec\right)
  \tag{1a}
\end{equation}

with

\begin{equation}
  \Eembed\!\left(\eps;\xvec\right)
  = \sum_{i=1}^{N} x_i F_\alpha\!\left(\rho_i\right)
  \tag{1b}
\end{equation}

\begin{equation}
  \rho_i
  = \sum_{j\neq i}^{N} x_j\,\rho_\beta\!\left(r_{ij}(\eps)\right)
  \tag{1c}
\end{equation}

\begin{equation}
  \Epair\!\left(\eps;\xvec\right)
  = \frac{1}{2}\sum_{i=1}^{N} x_i
    \sum_{j\neq i}^{N} x_j\,\phi_{\alpha\beta}\!\left(r_{ij}(\eps)\right)
  \tag{1d}
\end{equation}

where $r_{ij}$ is the distance between atoms $i$ and $j$, $F_\alpha$ is
the embedding energy to place atom $i$ of type $\alpha$ into the electron
cloud, $\rho_\beta$ is the contribution to the electron charge density
from atom $j$ of type $\beta$ at the location of atom $i$, and
$\phi_{\alpha\beta}$ is the pairwise potential energy between atom $i$
of type $\alpha$ and atom $j$ of type $\beta$.

\subsection*{Sensitivity analysis}

To connect the tangent stiffness of a nanostructure with its total EAM
energy, we expand the relaxed total energy about the undeformed state:

\begin{equation}
  \Etot\!\left(\eps;\xvec\right)
  = \Etot\!\left(0;\xvec\right)
    + \left(\frac{\partial \Etot}{\partial\eps}\right)_{\eps=0}\!\eps
    + \frac{1}{2}\left(\frac{\partial^2 \Etot}{\partial\eps^2}\right)_{\eps=0}\!\eps^2
    + \calO\!\left(\eps^3\right)
  \label{eq:Etaylor}
  \tag{2}
\end{equation}

The linear term arises from residual surface stress
$\tau_0(\xvec) \coloneqq \bigl(\partial \Etot/\partial\eps\bigr)_{\eps=0}$,
whereas the curvature term is the tangent stiffness
$\Keff(\xvec) \coloneqq \bigl(\partial^2 \Etot/\partial\eps^2\bigr)_{\eps=0}$.
A one-sided elastic strain energy
$\Etot(+\eps) - \Etot(0)$ mixes these two effects. To remove the
residual surface-stress bias, we define the symmetric energy-curvature
objective:

\begin{equation}
  \calJ\!\left(\xvec;\epszero\right)
  \coloneqq
  \frac{
    \Etot\!\left(+\epszero;\xvec\right)
    - 2\,\Etot\!\left(0;\xvec\right)
    + \Etot\!\left(-\epszero;\xvec\right)
  }{\epszero^{2}}
  \approx \Keff\!\left(\xvec\right)
  \label{eq:calJ}
  \tag{3}
\end{equation}

where $\calJ$ is a surface-stress-free measure of stiffness that
approximates the tangent stiffness as $\epszero \to 0$. We further
define the symmetric per-atom strain energy:

\begin{equation}
  \Ebar{i}\!\left(\epszero;\xvec\right)
  \coloneqq
  \frac{
    \Etilde{i}\!\left(+\epszero;\xvec\right)
    - 2\,\Etilde{i}\!\left(0;\xvec\right)
    + \Etilde{i}\!\left(-\epszero;\xvec\right)
  }{2}
  \label{eq:Ebar}
  \tag{4}
\end{equation}

Summing over atoms recovers the total symmetric strain energy:

\begin{equation}
  \sum_{i=1}^{N} \Ebar{i}\!\left(\epszero;\xvec\right)
  =
  \frac{
    \Etot\!\left(+\epszero;\xvec\right)
    - 2\,\Etot\!\left(0;\xvec\right)
    + \Etot\!\left(-\epszero;\xvec\right)
  }{2}
  = \tfrac{1}{2}\calJ\,\epszero^{2}
  \label{eq:sumEbar}
  \tag{5}
\end{equation}

To evaluate the contribution of each atom to the tangent stiffness, we
calculate the gradient (sensitivity) with respect to a design variable:

\begin{equation}
  \frac{\partial\calJ}{\partial x_k}
  = \frac{2}{\epszero^2}
    \sum_{i=1}^{N}
    \frac{\partial \Ebar{i}}{\partial x_k}
  \label{eq:dJdxk_a}
  \tag{6a}
\end{equation}

with

\begin{equation}
  \frac{\partial \Ebar{i}}{\partial x_k}
  = \frac{1}{2}\!\left(
      \frac{\partial \Etilde{i}(+\epszero;\xvec)}{\partial x_k}
      - 2\,\frac{\partial \Etilde{i}(0;\xvec)}{\partial x_k}
      + \frac{\partial \Etilde{i}(-\epszero;\xvec)}{\partial x_k}
    \right)
  \label{eq:dJdxk_b}
  \tag{6b}
\end{equation}

We write per-atom energy as

\begin{equation}
  \Etilde{i} = x_i A_i
  \label{eq:Etilde_a}
  \tag{7a}
\end{equation}

with

\begin{equation}
  A_i\!\left(\eps;\xvec\right)
  \coloneqq
  F_\alpha\!\left(\rho_i\right)
  + \frac{1}{2}
    \sum_{j\neq i}^{N}
    x_j\,\phi_{\alpha\beta}\!\left(r_{ij}(\eps)\right)
  \label{eq:Etilde_b}
  \tag{7b}
\end{equation}

We adopt the envelope theorem assumption. When differentiating with
respect to $x_k$, we treat the atomic positions as fixed, dropping the
implicit position derivatives $\partial r_{ij}/\partial x_k$ that arise
only via re-relaxation. Substituting into~\eqref{eq:dJdxk_a}:

\begin{equation}
  \frac{\partial\calJ}{\partial x_k}
  = \frac{2}{\epszero^2}
    \left(
      \Abar{k}
      + \sum_{i\neq k}^{N}
        x_i \frac{\partial \Abar{i}}{\partial x_k}
    \right)
  \label{eq:dJdxk_8a}
  \tag{8a}
\end{equation}

with

\begin{equation}
  \Abar{k}
  = \frac{1}{2}\!\left[
      A_k(+\epszero) - 2\,A_k(0) + A_k(-\epszero)
    \right]
  \label{eq:Abar_k}
  \tag{8b}
\end{equation}

\begin{equation}
  A_k(\eps)
  = F_\alpha\!\left(\rho_k(\eps)\right)
  + \frac{1}{2}
    \sum_{j\neq k}^{N}
    x_j\,\phi_{\alpha\beta}\!\left(r_{kj}(\eps)\right)
  \label{eq:Ak_eps}
  \tag{8c}
\end{equation}

and

\begin{align}
  \sum_{i\neq k}^{N} x_i \frac{\partial \Abar{i}}{\partial x_k}
  &= \sum_{i\neq k}^{N} x_i \frac{1}{2}\Bigl[
      F_\alpha'\!\left(\rho_i(+\epszero)\right)
      \frac{\partial\rho_i(+\epszero)}{\partial x_k}
      + \tfrac{1}{2}\phi_{\alpha\beta}\!\left(r_{ik}(+\epszero)\right)
    \notag\\
  &\quad
      - 2\Bigl\{
          F_\alpha'\!\left(\rho_i(0)\right)
          \frac{\partial\rho_i(0)}{\partial x_k}
          + \tfrac{1}{2}\phi_{\alpha\beta}\!\left(r_{ik}(0)\right)
        \Bigr\}
    \notag\\
  &\quad
      + F_\alpha'\!\left(\rho_i(-\epszero)\right)
        \frac{\partial\rho_i(-\epszero)}{\partial x_k}
      + \tfrac{1}{2}\phi_{\alpha\beta}\!\left(r_{ik}(-\epszero)\right)
    \Bigr]
  \approx \overline{A}_k^{\mathrm{pair}}
  \label{eq:sum_dAbar}
  \tag{8d}
\end{align}

Therefore,

\begin{equation}
  \frac{\partial\calJ}{\partial x_k}
  \approx
  \frac{2}{\epszero^2}
  \!\left(\Abar{k} + \overline{A}_k^{\mathrm{pair}}\right)
  = \frac{2}{\epszero^2}
    \!\left(
      \frac{\Ebar{k}}{x_k}
      + \frac{\overline{\widetilde{E}}_k^{\mathrm{pair}}}{x_k}
    \right)
  \approx
  \frac{4}{\epszero^2}
  \frac{\Ebar{k}}{x_k}
  \label{eq:dJdxk_final}
  \tag{8e}
\end{equation}

We use a symmetric ($\pm\eps$) measure of stiffness that cancels residual
surface stress and isolates curvature. With EAM, the stiffness can be
decomposed into a Cauchy-consistent part from the pair term and a
non-Cauchy correction part controlled by the curvature of the embedding
function. For aluminum under small symmetric strains, site densities
vary only slightly around their values at $\eps = 0$. Consequently, the
central-difference energy curvature is typically dominated by the pair
term, while the embedding term provides the smaller non-Cauchy
correction. Since LAMMPS does not output the pair and embedding energies
separately, our sensitivity analysis uses the symmetric strain energy
$\Ebar{k}$ without an explicit pair/embedding split. From~\eqref{eq:dJdxk_final},
$\Ebar{k}$ is the symmetric strain energy of atom $k$. The design
variable $x_k$ can be either 0 or 1. If atom $k$ is a real atom
($x_k = 1$), its sensitivity is approximated as its symmetric strain
energy (dropping the constant factor). If atom $k$ is a virtual atom
($x_k = 0$), its sensitivity is undefined (set to zero).

\subsection*{Sensitivity filtering}

The raw sensitivity value of atom $k$ is regularized by a weighted
neighborhood filter:

\begin{equation}
  \Shat_k
  = \frac{1}{\displaystyle\sum_{i=1}^{n}\Hhat_i}
    \sum_{i=1}^{n}
    \Hhat_i \frac{\Ebar{i}}{x_i}
  \label{eq:Shat}
  \tag{9a}
\end{equation}

\begin{equation}
  \Hhat_i
  = \rmin - \dist(k,i),
  \quad
  \bigl\{i \in \mathcal{N}(k) \mid \dist(k,i) \le \rmin\bigr\}
  \label{eq:Hhat}
  \tag{9b}
\end{equation}

where $\mathcal{N}(k)$ is the fixed neighborhood of atom $k$, $\Hhat_i$
is the weighting factor for atom $i$, $\dist(k,i)$ is the distance
between atoms $k$ and $i$, and $\rmin$ is the filter radius. The
sensitivity of a virtual atom is set to zero in the sensitivity
analysis. However, the filtered sensitivity of a virtual atom can be
non-zero when real atoms are present in its neighborhood, which allows
Nano-TO to determine which virtual atoms should be converted to real
atoms.

Unless otherwise noted, the filter radius is $\rmin = 10.325$~\AA,
which reaches the 13th FCC shell and therefore averages over the first 12
FCC shells (248 atoms total); the 13th shell has zero weight by
construction. In the reduced-scale finite-thickness nanocantilever study,
we instead use $\rmin = 4.05$~\AA\ to maintain a comparable relative
minimum feature size.

\subsection*{Nano-TO update scheme and convergence}

At each Nano-TO iteration, real atoms with the lowest filtered
sensitivity values are selected for removal, and virtual atoms with the
highest filtered sensitivity values are selected for insertion. For the
nanocantilever design problems, Nano-TO proceeds in two phases.
Optimization is initialized from a fully dense beam (mass ratio =
100\%). Independent trials are generated by applying a small random
displacement perturbation of magnitude $10^{-5}$~\AA\ in each Cartesian
direction before each energy minimization. In the mass-reduction phase,
each iteration converts 160 real active atoms to virtual atoms and
restores 80 virtual atoms to real atoms, for a net removal of 80 atoms,
until the target mass ratio is reached. A mass-conserving refinement
phase then converts 20 real active atoms to virtual atoms and restores
20 virtual atoms to real atoms per iteration. This second phase refines
the topology at a fixed mass ratio until convergence. For the
reduced-scale finite-thickness cantilevers, the mass-reduction phase
converts 8 real active atoms to virtual atoms and restores 4 virtual
atoms to real atoms. The mass-conserving refinement phase converts 4
real active atoms to virtual atoms and restores 4 virtual atoms per
iteration.

Convergence is declared when the optimization enters a period-two cycle:
the sets of atoms converted at iteration $n$ are exactly reversed at
iteration $n+1$, indicating that no further net improvement is achieved
under the current update rule. For the nanopillars, only the
mass-conserving refinement phase is applied, converting 80 real active
atoms to virtual atoms and restoring 80 virtual atoms per iteration.
Convergence is not achieved after 3{,}000 iterations; optimization is
terminated due to compute budget.

\subsection*{Generation of designs based on Gaussian random fields}

We generate large batches of synthetic beam layouts by sampling
continuous Gaussian random fields (GRFs) on a rectangular grid and
converting them to binary designs encoding real atoms (1) or virtual
atoms (0). We use a spectral (Fourier) method. Complex white noise in
frequency space is filtered by a Gaussian-shaped power spectrum
(squared-exponential kernel with correlation parameters $l_x$, $l_y$)
and mapped to real space by the inverse discrete Fourier transform.

Fields are generated on an oversampled grid of size $450\times75$ and
then center-cropped to the final $300\times50$ domain to suppress
periodic artifacts. A target mass ratio is imposed by rank-order
thresholding. After thresholding, mid-plane symmetry is enforced by
reflecting the image. Each binary layout is mapped to an atomistic
thickness-periodic nanocantilever model in which each image pixel
corresponds to one FCC column in the periodic thickness direction. A
candidate is rejected if it contains disconnected floating regions, if no
real pixels touch the clamped boundary, or if no real pixels are present
at the loaded free tip.

\subsection*{Conditional denoising diffusion probabilistic models}

We construct conditional denoising diffusion probabilistic models
(c-DDPMs) to synthesize thickness-periodic nanocantilever layouts with
target properties. Each design is encoded as a $100\times300$ binary
image, where a value of 1 denotes a real-atom column and 0 denotes a
virtual-atom column. Since mirror symmetry about the beam mid-plane is
enforced, only one half of each image is modeled explicitly during
training, and the full design is reconstructed by reflection before
atomistic evaluation. Training uses the symmetric half-image,
zero-padded to $64\times304$, and rescales inputs to $[-1,1]$ by

\begin{equation}
  \widetilde{\xvec}_0 = 2\xvec_0 - 1
  \label{eq:rescale}
  \tag{10}
\end{equation}

The conditioning variable is task-dependent. For Gaussian-DDPM, the
condition is the scalar normalized bending stiffness $k$. For TO-DDPM,
the conditioning vector $\cvec = (k,m) \in \mathbb{R}^2$, where $k$ and
$m$ denote normalized bending stiffness and mass ratio, respectively.
Each conditioning component is linearly scaled to $[-1,1]$ by

\begin{equation}
  \widetilde{y}
  = 2\,\frac{y - y_{\min}}{y_{\max} - y_{\min}} - 1
  \label{eq:yscale}
  \tag{11}
\end{equation}

We adopt the standard DDPM forward noising process~\cite{Ho2020}:

\begin{equation}
  q\!\left(\xvec_t \mid \xvec_{t-1}\right)
  = \calN\!\left(\sqrt{1-\beta_t}\,\xvec_{t-1},\;\beta_t\mathbf{I}\right)
  \label{eq:DDPM_forward}
  \tag{12}
\end{equation}

with $\alpha_t = 1 - \beta_t$ and
$\bar\alpha_t = \prod_{s=1}^{t}\alpha_s$. The corresponding closed-form
reparameterization is

\begin{equation}
  \xvec_t
  = \sqrt{\bar\alpha_t}\,\xvec_0
    + \sqrt{1 - \bar\alpha_t}\,\bm{\varepsilon}
  \label{eq:reparam_a}
  \tag{13a}
\end{equation}

where

\begin{equation}
  \bm{\varepsilon} \sim \calN(\mathbf{0}, \mathbf{I})
  \label{eq:reparam_b}
  \tag{13b}
\end{equation}

using a linear $\beta_t$ schedule with $T = 1{,}000$ diffusion steps.

The denoiser
$\varepsilon_\theta\!\left(\xvec_t, t, \cvec\right)$
is a U-Net backbone~\cite{Ronneberger2015} augmented with
cross-attention~\cite{Vaswani2017}. A learned embedding of the target
properties, produced by a multilayer perceptron (MLP), modulates all
stages of the network, following the cross-attention conditioning used
in modern diffusion models~\cite{Rombach2022}. We train the network with
the standard predict-the-noise parameterization of DDPM~\cite{Ho2020}
with the mean-squared-error loss:

\begin{equation}
  \mathcal{L}(\theta,\phi)
  = \mathbb{E}_{\xvec_0,\,t,\,\cvec,\,\bm{\varepsilon}}
    \!\left[
      \bigl\|
        \bm{\varepsilon}
        - \varepsilon_\theta\!\left(\xvec_t, t, \cvec\right)
      \bigr\|_2^2
    \right]
  \label{eq:loss}
  \tag{14}
\end{equation}

With per-dimension dropout $p = 0.1$ and conditioning variables $D = 2$ (as in TO-DDPM), the probability of
fully unconditioned training samples is $p^{D} = 0.01$ and fully
conditioned samples occur with $(1-p)^{D} = 0.81$. Therefore, the
probability of partially dropped samples is 0.18. This mixture trains
the model to handle unconditional, partially conditional, and fully
conditional inputs with a single set of weights, as advocated by
classifier-free guidance (CFG)~\cite{Ho2022CFG}.

Our null condition is the all-zero vector $\mathbf{0}$.
The conditioning MLP maps both real conditions $\cvec$ and the null
$\mathbf{0}$ to embeddings that drive cross-attention in the U-Net. At
inference, we run two forward passes per timestep: one with the null
condition and one with the target condition. Let
$\bm{\varepsilon}^u = \varepsilon_\theta(\xvec_t, t, \mathbf{0})$ and
$\bm{\varepsilon}^c = \varepsilon_\theta(\xvec_t, t, \cvec)$. The
guided noise estimate is

\begin{equation}
  \widehat{\bm{\varepsilon}}
  = \bm{\varepsilon}^u
    + w\!\left(\bm{\varepsilon}^c - \bm{\varepsilon}^u\right)
  \label{eq:CFG}
  \tag{15}
\end{equation}

where $w$ is the guidance strength. The reverse-diffusion update then
uses $\widehat{\bm{\varepsilon}}$ in the DDPM posterior. Intuitively,
the guidance strength trades off fidelity to the condition (larger $w$)
against sample diversity (smaller $w$). Increasing the guidance strength
typically sharpens compliance with target properties but can reduce
variety or introduce artifacts if pushed too far~\cite{Ho2022CFG}. All
c-DDPMs are trained on an NVIDIA RTX A6000 GPU.

\subsection*{Task-specific conditioning and guidance selection}

For the Gaussian-DDPM benchmark, the GRF dataset is labeled only by
normalized bending stiffness. High-stiffness sampling is performed at
the upper-bound condition $k = 1.0$. Classifier-free guidance strengths
$w \in \{1, 3, 5, 7\}$ are evaluated by generating 1{,}600 samples at
each $w$, converting each sample to an atomistic model, and measuring
its normalized bending stiffness. The setting $w = 3$ is used for the
Gaussian-DDPM results as it provides a strong trade-off between property
targeting and sample diversity (see Supplementary Notes A.7).

For TO-DDPM, each sample is labeled by both normalized bending stiffness
$k$ and mass ratio $m$. Mass-ratio labels are linearly mapped, and
$m = -1.0$ corresponds to the target mass ratio of 59.60\%. Since the
maximum achievable stiffness depends on the mass ratio, setting $k = 1.0$
together with $m = -1.0$ imposes an unattainable target combination. We
therefore evaluate stiffness conditions $k \in \{0.0, -0.2, -0.4, -0.6,
-0.8, -1.0\}$ together with guidance strengths $w \in \{1, 3, 5, 7\}$,
while fixing $m = -1.0$. For each $(k, w)$ pair, 1{,}600 samples are
generated and evaluated. The setting $k = -0.6$ and $w = 1$ is used for
the TO-DDPM results (see Supplementary Notes A.8).

\subsection*{Data availability}

All data used in this study were generated directly from the code.

\subsection*{Code availability}

The code used in this study is publicly available at:
\url{https://github.com/chunteh/Diffusion-Nano-TO}

\clearpage
\section*{Figures}

\begin{figure}[H]
  \centering
  \includegraphics[width=\textwidth]{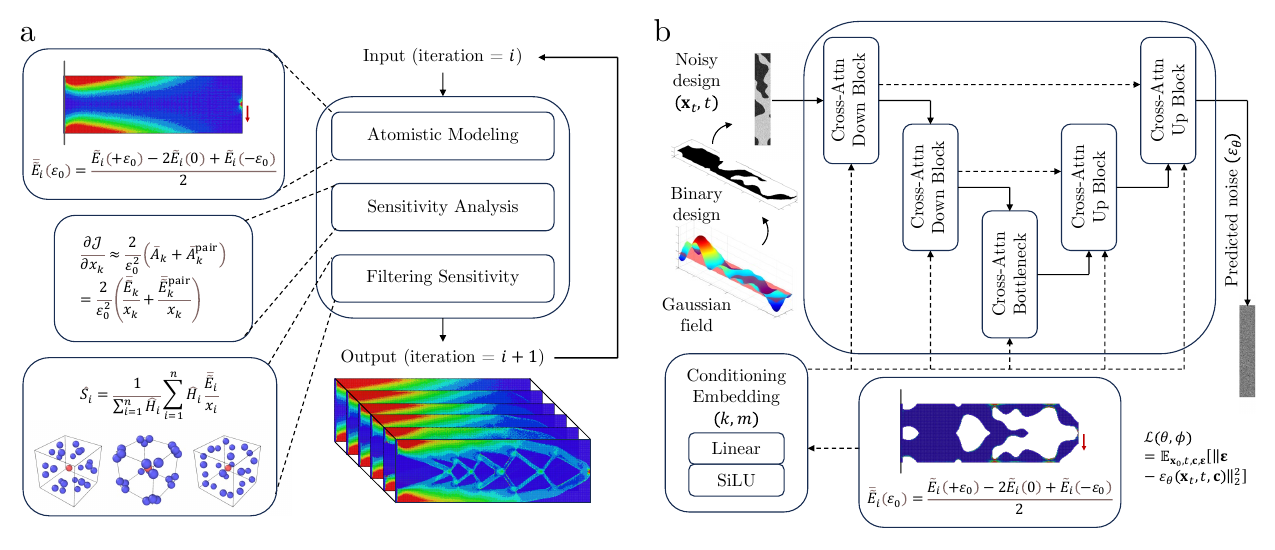}
  \caption{\textbf{Nano-TO and c-DDPM frameworks.}
    \textbf{a}, Nano-TO workflow for designing nanostructures by
    iteratively adding and removing atoms. Each iteration starts from
    the current design configuration and undergoes: (1) atomistic
    modeling using the embedded-atom method, (2) sensitivity analysis
    to estimate each atom's contribution to mechanical properties, and
    (3) sensitivity filtering to smooth the sensitivity analysis values.
    The algorithm then updates the design and repeats until convergence.
    \textbf{b}, c-DDPM framework. The property of interest (e.g.,
    bending stiffness, mass ratio) is mapped into a conditioning embedding,
    which guides a U-Net denoiser through cross-attention layers.
    Nanobeams are represented as images (binary), and the network learns
    to iteratively remove noise from perturbed samples to generate new
    designs matching target properties.}
  \label{fig:1}
\end{figure}

\begin{figure}[H]
  \centering
  \includegraphics[width=0.92\textwidth]{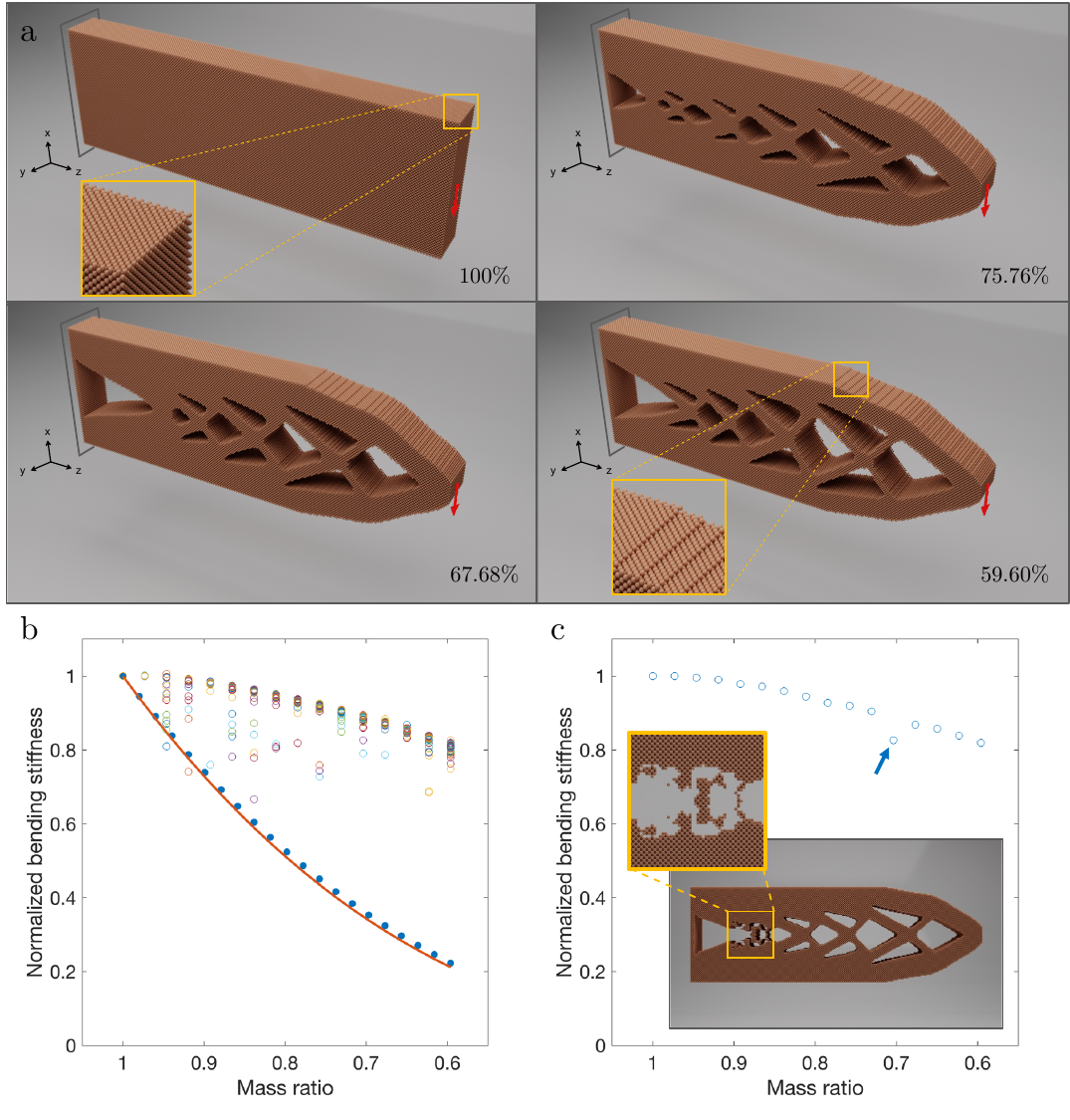}
  \caption{\textbf{Nano-TO design of thickness-periodic
    nanocantilevers.}
    \textbf{a}, Initial design (100\%) and optimized designs at
    different mass ratios (75.76\%, 67.68\%, and 59.60\%). The gray
    block represents the clamped support; the red arrow marks the
    applied vertical displacement at the free end. Models are rendered
    with three periodic images in the thickness direction. Insets show
    local atomic arrangements, illustrating how atoms are selectively
    removed to form truss-like motifs with multiple cross-braces.
    \textbf{b}, Normalized bending stiffness versus mass ratio. Colored
    circles: results from 64 independent trials at each mass ratio.
    Blue dots: height-scaled reference beams. Red curve:
    Euler--Bernoulli estimate.
    \textbf{c}, Example optimization trajectory showing a transient
    drop in stiffness (arrow) caused by a pattern transition that
    creates disconnected ``floating'' atoms (inset).}
  \label{fig:2}
\end{figure}

\begin{figure}[H]
  \centering
  \includegraphics[width=0.92\textwidth]{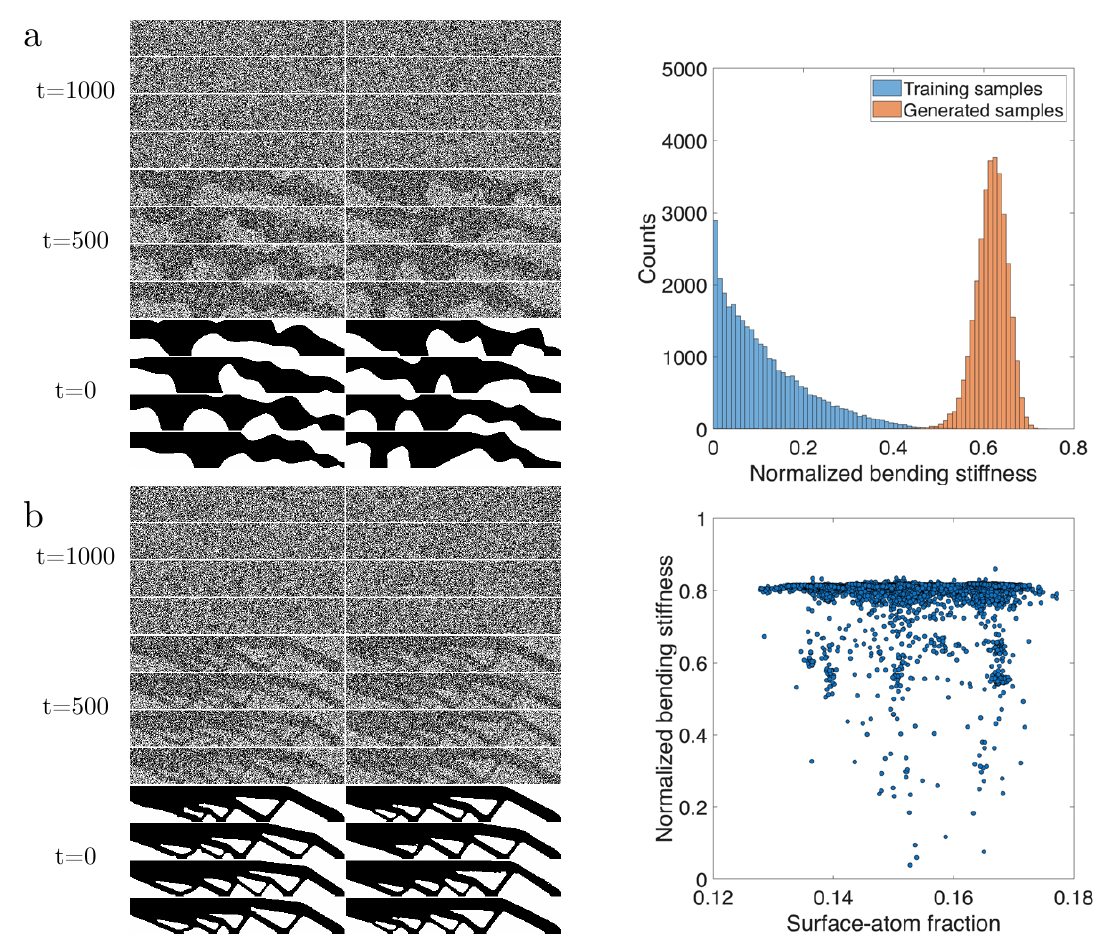}
  \caption{\textbf{c-DDPM denoising trajectories and performance.}
    \textbf{a}, Gaussian-DDPM. The left panel shows denoising snapshots
    at $t = 1{,}000$, 500, and 0, in which random noise is gradually
    refined into smooth, curved motifs typical of GRF layouts (bottom
    row). The right panel shows the histogram of normalized bending
    stiffness for the training samples (blue) versus the generated
    designs (red). \textbf{b}, TO-DDPM. The left panel shows snapshots
    of denoising at $t = 1{,}000$, 500, and 0, in which random noise
    gradually organizes into truss-like motifs similar to Nano-TO
    designs. The right panel plots normalized bending stiffness versus
    surface-atom fraction for the generated designs.}
  \label{fig:3}
\end{figure}

\begin{figure}[H]
  \centering
  \includegraphics[width=0.92\textwidth]{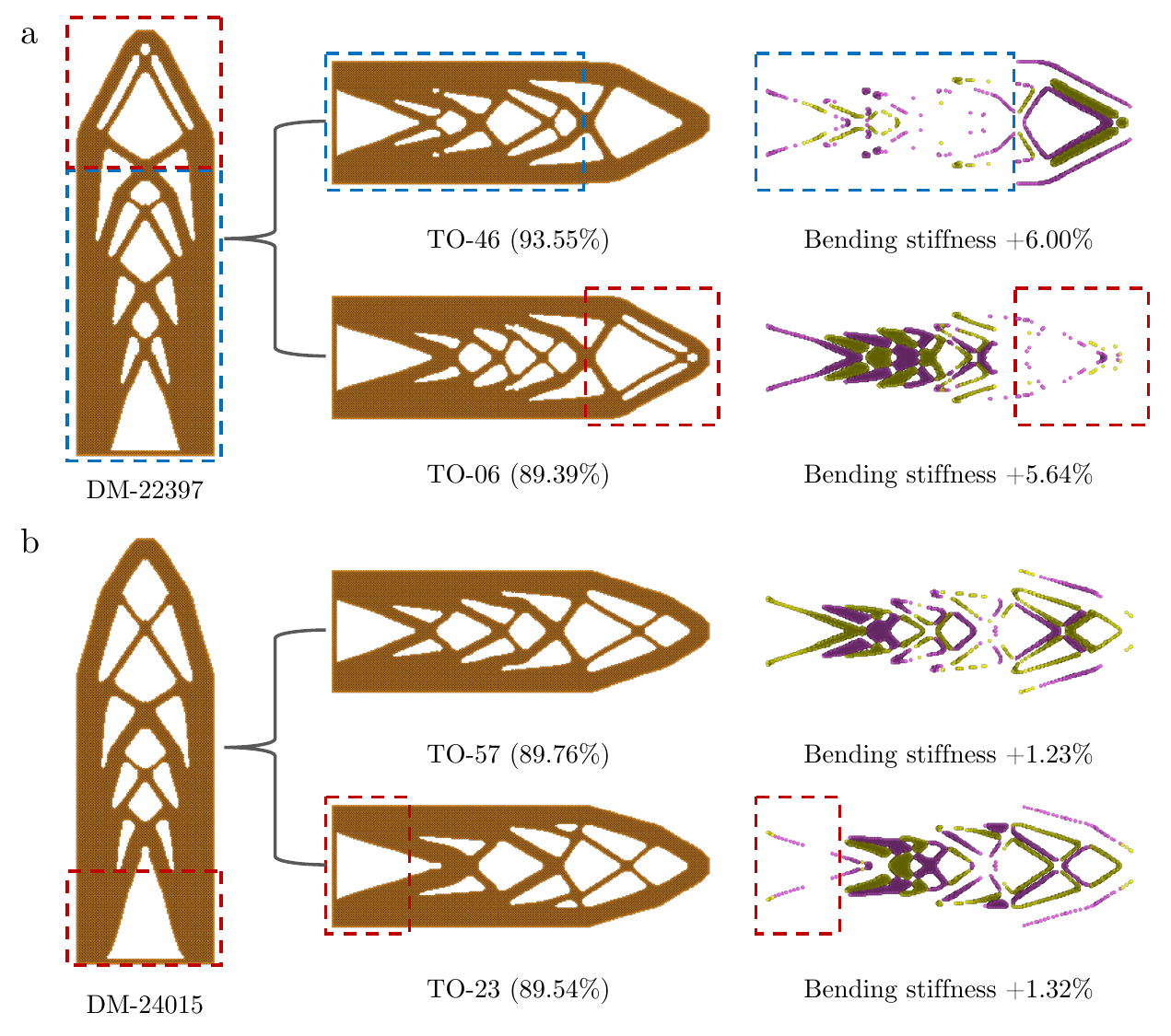}
  \caption{\textbf{Diffusion as a recombination and local refinement
    operator on a near-optimal manifold.}
    In both panels, the overlay uses yellow to mark atoms present only in
    the generated design and magenta for atoms present only in the
    training sample; stiffness gains are relative to the respective
    training sample.
    \textbf{a},~DM-22397 and its two nearest training samples.
    The blue dashed box marks the region of DM-22397 that closely
    resembles TO-46; the red dashed box marks the region that resembles
    TO-06.  In each overlay, the corresponding boxed region contains
    noticeably fewer atomic differences, indicating that the diffusion
    model has recombined sub-structures from two distinct training
    samples into a single higher-performing design.
    \textbf{b},~DM-24015 and its two nearest training samples.
    Only the red-boxed region bears clear resemblance to TO-23; the
    global topology differs substantially from both nearest neighbors.
    This example shows that TO-DDPM does not merely replicate or splice
    existing training samples but can synthesize topologies informed by
    the full learned distribution.}
  \label{fig:4}
\end{figure}

\begin{figure}[H]
  \centering
  \includegraphics[width=0.92\textwidth]{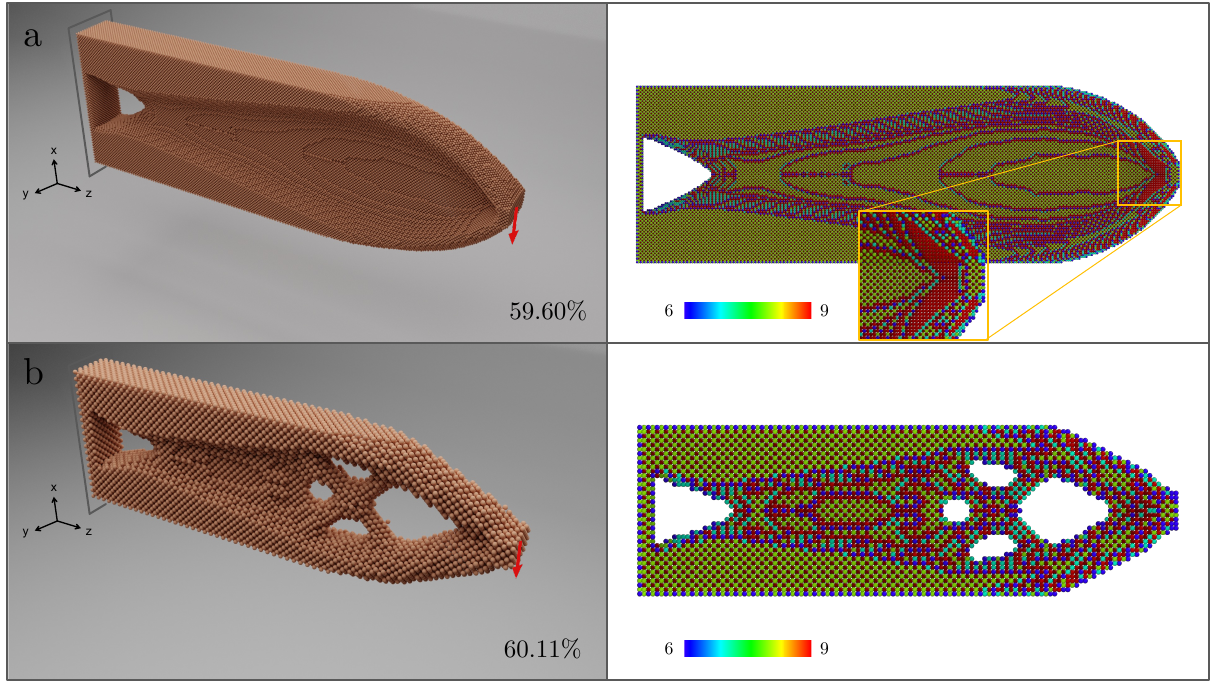}
  \caption{\textbf{Size-dependent optimized designs of finite-thickness
    nanocantilevers.}
    \textbf{a}, Optimized finite-thickness design at a mass ratio of
    59.60\%, colored by coordination number (6--9). The yellow box
    and inset show that Nano-TO preferentially exposes
    \{111\} facets (coordination number~9, red), the stiffest
    FCC surfaces, to locally maximize bending stiffness.
    \textbf{b},
    Optimized finite-thickness design at a mass ratio of 60.11\% with
    all dimensions reduced to approximately 40\%, colored by
    coordination number (6--9). As dimensions shrink, the continuous wall
    in panel~\textbf{a} becomes only a few atomic layers thick, and its
    ability to carry transverse shear degrades. The optimized design
    then redirects shear into inclined cross-braces, producing a
    truss-like configuration.}
  \label{fig:5}
\end{figure}

\begin{figure}[H]
  \centering
  \includegraphics[width=0.92\textwidth]{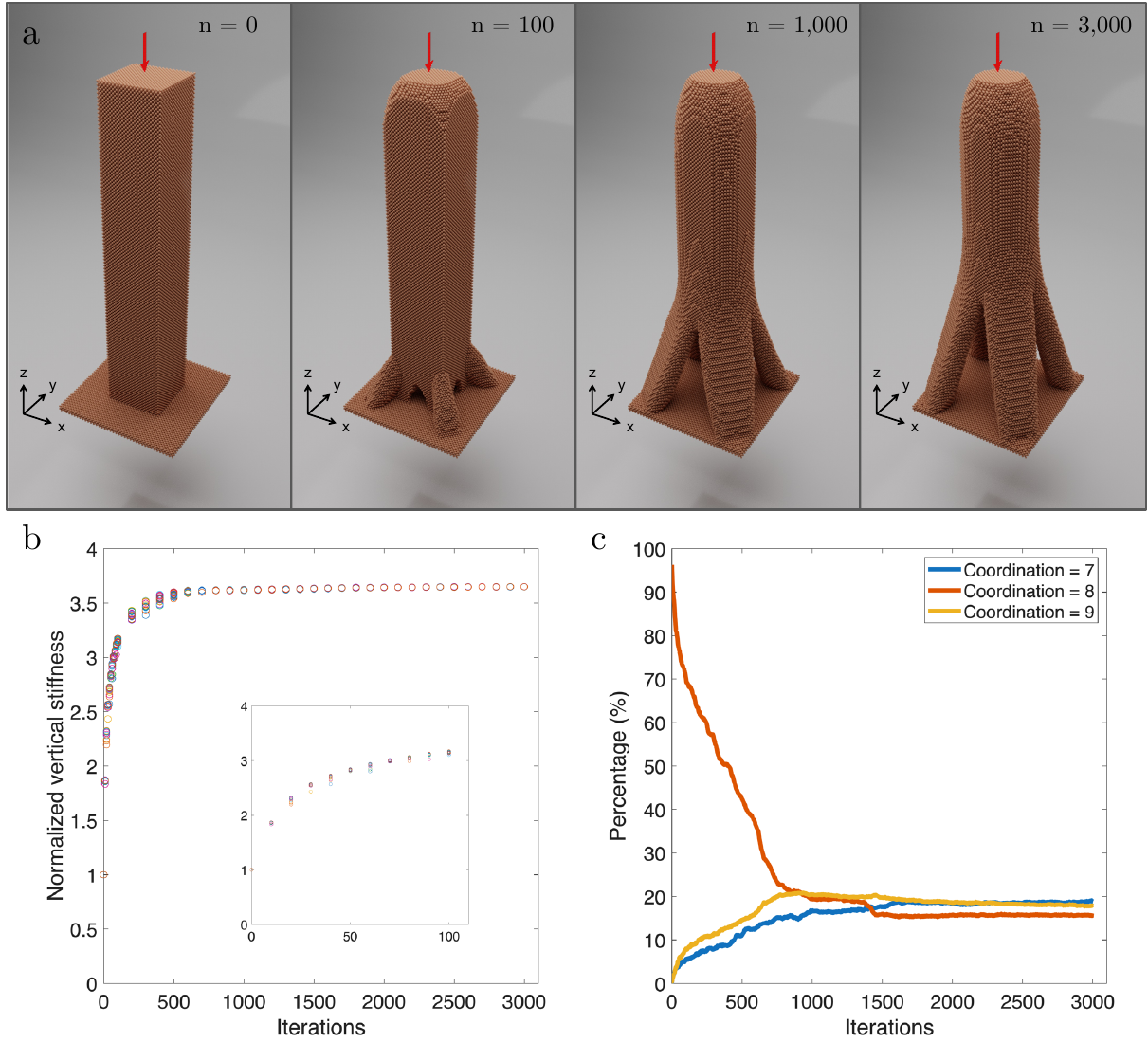}
  \caption{\textbf{Nano-TO design of nanopillars.}
    \textbf{a}, Initial design and optimized designs at different
    iterations (100, 1{,}000, and 3{,}000). The red arrow marks the
    applied vertical displacement at the center of the top surface.
    \textbf{b}, Normalized vertical stiffness versus iterations from 16
    independent trials at each iteration, showing rapid gains in the
    first 100 iterations (inset) and an apparent plateau at a
    stiffness 3.65 times the initial value.
    \textbf{c}, Evolution of coordination numbers 7, 8, and 9
    throughout the design process. These roughly correspond to
    $\{110\}$-, $\{100\}$-, and $\{111\}$-like surfaces, respectively,
    demonstrating how the optimized design balances surface orientations
    to enhance stiffness.}
  \label{fig:6}
\end{figure}

\clearpage
\bibliographystyle{unsrtnat}
\bibliography{references}

\begin{thebibliography}{49}
\providecommand{\natexlab}[1]{#1}
\providecommand{\url}[1]{\texttt{#1}}
\expandafter\ifx\csname urlstyle\endcsname\relax
  \providecommand{\doi}[1]{doi: #1}\else
  \providecommand{\doi}{doi: \begingroup \urlstyle{rm}\Url}\fi

\bibitem[Rugar et~al.(2004)Rugar, Budakian, Mamin, and Chui]{Rugar2004}
D.~Rugar, R.~Budakian, H.~Mamin, and B.~Chui.
\newblock Single spin detection by magnetic resonance force microscopy.
\newblock \emph{Nature}, 430:\penalty0 329--332, 2004.

\bibitem[Ekinci and Roukes(2005)]{Ekinci2005}
K.~L. Ekinci and M.~L. Roukes.
\newblock Nanoelectromechanical systems.
\newblock \emph{Review of Scientific Instruments}, 76:\penalty0 061101, 2005.

\bibitem[Trimble et~al.(2003)Trimble, Cammarata, and Sieradzki]{Trimble2003}
T.~Trimble, R.~Cammarata, and K.~Sieradzki.
\newblock The stability of fcc (1\,1\,1) metal surfaces.
\newblock \emph{Surface Science}, 531:\penalty0 8--20, 2003.

\bibitem[Deng and Sansoz(2009)]{Deng2009}
C.~Deng and F.~Sansoz.
\newblock Near-ideal strength in gold nanowires achieved through
  microstructural design.
\newblock \emph{ACS Nano}, 3:\penalty0 3001--3008, 2009.

\bibitem[Shenoy(2005)]{Shenoy2005}
V.~B. Shenoy.
\newblock Atomistic calculations of elastic properties of metallic {FCC}
  crystal surfaces.
\newblock \emph{Physical Review B}, 71:\penalty0 094104, 2005.

\bibitem[Zhang et~al.(2008)Zhang, Luo, and Chan]{Zhang2008}
T.-Y. Zhang, M.~Luo, and W.~K. Chan.
\newblock Size-dependent surface stress, surface stiffness, and {Young's}
  modulus of hexagonal prism [111] $\beta$-{SiC} nanowires.
\newblock \emph{Journal of Applied Physics}, 103:\penalty0 104308, 2008.

\bibitem[Wang and Li(2008)]{Wang2008}
G.~Wang and X.~Li.
\newblock Predicting {Young's} modulus of nanowires from first-principles
  calculations on their surface and bulk materials.
\newblock \emph{Journal of Applied Physics}, 104:\penalty0 113517, 2008.

\bibitem[Zhu et~al.(2012)]{Zhu2012}
Y.~Zhu et~al.
\newblock Size effects on elasticity, yielding, and fracture of silver
  nanowires: in situ experiments.
\newblock \emph{Physical Review B}, 85:\penalty0 045443, 2012.

\bibitem[Miller and Shenoy(2000)]{Miller2000}
R.~E. Miller and V.~B. Shenoy.
\newblock Size-dependent elastic properties of nanosized structural elements.
\newblock \emph{Nanotechnology}, 11:\penalty0 139--147, 2000.

\bibitem[Cuenot et~al.(2004)Cuenot, Fr{\'e}tigny, Demoustier-Champagne, and
  Nysten]{Cuenot2004}
S.~Cuenot, C.~Fr{\'e}tigny, S.~Demoustier-Champagne, and B.~Nysten.
\newblock Surface tension effect on the mechanical properties of nanomaterials
  measured by atomic force microscopy.
\newblock \emph{Physical Review B}, 69:\penalty0 165410, 2004.

\bibitem[Bends{\o}e and Kikuchi(1988)]{Bendsoe1988}
M.~P. Bends{\o}e and N.~Kikuchi.
\newblock Generating optimal topologies in structural design using a
  homogenization method.
\newblock \emph{Computer Methods in Applied Mechanics and Engineering},
  71:\penalty0 197--224, 1988.

\bibitem[Bends{\o}e(1989)]{Bendsoe1989}
M.~P. Bends{\o}e.
\newblock Optimal shape design as a material distribution problem.
\newblock \emph{Structural Optimization}, 1:\penalty0 193--202, 1989.

\bibitem[Bends{\o}e and Sigmund(2003)]{Bendsoe2013}
M.~P. Bends{\o}e and O.~Sigmund.
\newblock \emph{Topology Optimization: Theory, Methods, and Applications}.
\newblock Springer, 2003.

\bibitem[Eschenauer and Olhoff(2001)]{Eschenauer2001}
H.~A. Eschenauer and N.~Olhoff.
\newblock Topology optimization of continuum structures: a review.
\newblock \emph{Applied Mechanics Reviews}, 54:\penalty0 331--390, 2001.

\bibitem[Sigmund and Maute(2013)]{Sigmund2013}
O.~Sigmund and K.~Maute.
\newblock Topology optimization approaches: A comparative review.
\newblock \emph{Structural and Multidisciplinary Optimization}, 48:\penalty0
  1031--1055, 2013.

\bibitem[Aage et~al.(2017)Aage, Andreassen, Lazarov, and Sigmund]{Aage2017}
N.~Aage, E.~Andreassen, B.~S. Lazarov, and O.~Sigmund.
\newblock Giga-voxel computational morphogenesis for structural design.
\newblock \emph{Nature}, 550:\penalty0 84--86, 2017.

\bibitem[Andreassen et~al.(2011)Andreassen, Clausen, Schevenels, Lazarov, and
  Sigmund]{Andreassen2011}
E.~Andreassen, A.~Clausen, M.~Schevenels, B.~S. Lazarov, and O.~Sigmund.
\newblock Efficient topology optimization in {MATLAB} using 88 lines of code.
\newblock \emph{Structural and Multidisciplinary Optimization}, 43:\penalty0
  1--16, 2011.

\bibitem[Liu and Tovar(2014)]{Liu2014}
K.~Liu and A.~Tovar.
\newblock An efficient {3D} topology optimization code written in {Matlab}.
\newblock \emph{Structural and Multidisciplinary Optimization}, 50:\penalty0
  1175--1196, 2014.

\bibitem[Gurtin and Ian~Murdoch(1975)]{Gurtin1975}
M.~E. Gurtin and A.~Ian~Murdoch.
\newblock A continuum theory of elastic material surfaces.
\newblock \emph{Archive for Rational Mechanics and Analysis}, 57:\penalty0
  291--323, 1975.

\bibitem[Zhu et~al.(2017)Zhu, Wei, and Guo]{Zhu2017}
Y.~Zhu, Y.~Wei, and X.~Guo.
\newblock {Gurtin--Murdoch} surface elasticity theory revisit: an orbital-free
  density functional theory perspective.
\newblock \emph{Journal of the Mechanics and Physics of Solids}, 109:\penalty0
  178--197, 2017.

\bibitem[Nanthakumar et~al.(2015)Nanthakumar, Valizadeh, Park, and
  Rabczuk]{Nanthakumar2015}
S.~Nanthakumar, N.~Valizadeh, H.~S. Park, and T.~Rabczuk.
\newblock Surface effects on shape and topology optimization of nanostructures.
\newblock \emph{Computational Mechanics}, 56:\penalty0 97--112, 2015.

\bibitem[Lam et~al.(2003)Lam, Yang, Chong, Wang, and Tong]{Lam2003}
D.~C. Lam, F.~Yang, A.~Chong, J.~Wang, and P.~Tong.
\newblock Experiments and theory in strain gradient elasticity.
\newblock \emph{Journal of the Mechanics and Physics of Solids}, 51:\penalty0
  1477--1508, 2003.

\bibitem[Mindlin(1965)]{Mindlin1965}
R.~D. Mindlin.
\newblock Second gradient of strain and surface-tension in linear elasticity.
\newblock \emph{International Journal of Solids and Structures}, 1:\penalty0
  417--438, 1965.

\bibitem[Chen et~al.(2020)Chen, Chrzan, and Gu]{Chen2020NanoTO}
C.-T. Chen, D.~C. Chrzan, and G.~X. Gu.
\newblock Nano-topology optimization for materials design with atom-by-atom
  control.
\newblock \emph{Nature Communications}, 11:\penalty0 3745, 2020.

\bibitem[Sigmund(2007)]{Sigmund2007}
O.~Sigmund.
\newblock Morphology-based black and white filters for topology optimization.
\newblock \emph{Structural and Multidisciplinary Optimization}, 33:\penalty0
  401--424, 2007.

\bibitem[Guest et~al.(2004)Guest, Pr{\'e}vost, and Belytschko]{Guest2004}
J.~K. Guest, J.~H. Pr{\'e}vost, and T.~Belytschko.
\newblock Achieving minimum length scale in topology optimization using nodal
  design variables and projection functions.
\newblock \emph{International Journal for Numerical Methods in Engineering},
  61:\penalty0 238--254, 2004.

\bibitem[Lazarov and Sigmund(2011)]{Lazarov2011}
B.~S. Lazarov and O.~Sigmund.
\newblock Filters in topology optimization based on {Helmholtz}-type
  differential equations.
\newblock \emph{International Journal for Numerical Methods in Engineering},
  86:\penalty0 765--781, 2011.

\bibitem[Sohl-Dickstein et~al.(2015)Sohl-Dickstein, Weiss, Maheswaranathan, and
  Ganguli]{SohlDickstein2015}
J.~Sohl-Dickstein, E.~Weiss, N.~Maheswaranathan, and S.~Ganguli.
\newblock Deep unsupervised learning using nonequilibrium thermodynamics.
\newblock In \emph{Proceedings of the 32nd International Conference on Machine
  Learning}, volume~37 of \emph{Proceedings of Machine Learning Research},
  pages 2256--2265. PMLR, 2015.

\bibitem[Ho et~al.(2020)Ho, Jain, and Abbeel]{Ho2020}
J.~Ho, A.~Jain, and P.~Abbeel.
\newblock Denoising diffusion probabilistic models.
\newblock \emph{Advances in Neural Information Processing Systems},
  33:\penalty0 6840--6851, 2020.

\bibitem[Song et~al.(2021)Song, Sohl-Dickstein, Kingma, Kumar, Ermon, and
  Poole]{Song2021}
Y.~Song, J.~Sohl-Dickstein, D.~P. Kingma, A.~Kumar, S.~Ermon, and B.~Poole.
\newblock Score-based generative modeling through stochastic differential
  equations.
\newblock In \emph{International Conference on Learning Representations}, 2021.

\bibitem[Chen and Gu(2020)]{Chen2019backprop}
C.-T. Chen and G.~X. Gu.
\newblock Generative deep neural networks for inverse materials design using
  backpropagation and active learning.
\newblock \emph{Advanced Science}, 7:\penalty0 1902607, 2020.

\bibitem[Kang et~al.(2024)Kang, Song, Kang, Bae, and Ryu]{Kang2024}
S.~Kang, H.~Song, H.~S. Kang, B.-S. Bae, and S.~Ryu.
\newblock Customizable metamaterial design for desired strain-dependent
  {Poisson's} ratio using constrained generative inverse design network.
\newblock \emph{Materials \& Design}, 247:\penalty0 113377, 2024.

\bibitem[S{\'a}nchez-Lengeling and Aspuru-Guzik(2018)]{SanchezLengeling2018}
B.~S{\'a}nchez-Lengeling and A.~Aspuru-Guzik.
\newblock Inverse molecular design using machine learning: Generative models
  for matter engineering.
\newblock \emph{Science}, 361:\penalty0 360--365, 2018.

\bibitem[Zheng et~al.(2023)Zheng, Karapiperis, Kumar, and
  Kochmann]{Zheng2023truss}
L.~Zheng, K.~Karapiperis, S.~Kumar, and D.~M. Kochmann.
\newblock Unifying the design space and optimizing linear and nonlinear truss
  metamaterials by generative modeling.
\newblock \emph{Nature Communications}, 14:\penalty0 7563, 2023.

\bibitem[Mao et~al.(2020)Mao, He, and Zhao]{Mao2020GAN}
Y.~Mao, Q.~He, and X.~Zhao.
\newblock Designing complex architectured materials with generative adversarial
  networks.
\newblock \emph{Science Advances}, 6:\penalty0 eaaz4169, 2020.

\bibitem[Bastek and Kochmann(2023)]{Bastek2023}
J.-H. Bastek and D.~M. Kochmann.
\newblock Inverse design of nonlinear mechanical metamaterials via video
  denoising diffusion models.
\newblock \emph{Nature Machine Intelligence}, 5:\penalty0 1466--1475, 2023.

\bibitem[Li et~al.(2026)Li, Wang, Jin, Zong, Zhu, Wang, Wang, Yang, Yin, and
  Wei]{Li2026}
E.~Li, Y.~Wang, L.~Jin, Z.~Zong, E.~Zhu, B.~Wang, Q.~Wang, Z.~Yang, W.-Y. Yin,
  and Z.~Wei.
\newblock Current-diffusion model for metasurface structure discoveries with
  spatial-frequency dynamics.
\newblock \emph{Nature Machine Intelligence}, 8:\penalty0 59--69, 2026.

\bibitem[Mishin et~al.(1999)Mishin, Farkas, Mehl, and
  Papaconstantopoulos]{Mishin1999}
Y.~Mishin, D.~Farkas, M.~Mehl, and D.~Papaconstantopoulos.
\newblock Interatomic potentials for monoatomic metals from experimental data
  and ab initio calculations.
\newblock \emph{Physical Review B}, 59:\penalty0 3393--3407, 1999.

\bibitem[Daw and Baskes(1984)]{Daw1984}
M.~S. Daw and M.~I. Baskes.
\newblock Embedded-atom method: Derivation and application to impurities,
  surfaces, and other defects in metals.
\newblock \emph{Physical Review B}, 29:\penalty0 6443--6453, 1984.

\bibitem[Dhariwal and Nichol(2021)]{Dhariwal2021}
P.~Dhariwal and A.~Nichol.
\newblock Diffusion models beat {GAN}s on image synthesis.
\newblock In \emph{Advances in Neural Information Processing Systems},
  volume~34, pages 8780--8794, 2021.

\bibitem[Ho and Salimans(2021)]{Ho2022CFG}
J.~Ho and T.~Salimans.
\newblock Classifier-free diffusion guidance.
\newblock In \emph{Workshop on Deep Generative Models and Downstream
  Applications at the 35th Conference on Neural Information Processing Systems
  (NeurIPS 2021)}, 2021.

\bibitem[Rieser and Zimmermann(2023)]{Rieser2023}
J.~Rieser and M.~Zimmermann.
\newblock Towards closed-walled designs in topology optimization using
  selective penalization.
\newblock \emph{Structural and Multidisciplinary Optimization}, 66:\penalty0
  158, 2023.

\bibitem[Sigmund et~al.(2016)Sigmund, Aage, and Andreassen]{Sigmund2016}
O.~Sigmund, N.~Aage, and E.~Andreassen.
\newblock On the (non-)optimality of {Michell} structures.
\newblock \emph{Structural and Multidisciplinary Optimization}, 54:\penalty0
  361--373, 2016.

\bibitem[Plimpton(1995)]{Plimpton1995}
S.~Plimpton.
\newblock Fast parallel algorithms for short-range molecular dynamics.
\newblock \emph{Journal of Computational Physics}, 117:\penalty0 1--19, 1995.

\bibitem[Thompson et~al.(2022)Thompson, Aktulga, Berger, Bolintineanu, Brown,
  Crozier, {in 't Veld}, Kohlmeyer, Moore, Nguyen, Shan, Stevens, Tranchida,
  Trott, and Plimpton]{Thompson2022}
A.~P. Thompson, H.~M. Aktulga, R.~Berger, D.~S. Bolintineanu, W.~M. Brown,
  P.~S. Crozier, P.~J. {in 't Veld}, A.~Kohlmeyer, S.~G. Moore, T.~D. Nguyen,
  R.~Shan, M.~J. Stevens, J.~Tranchida, C.~Trott, and S.~J. Plimpton.
\newblock {LAMMPS} --- a flexible simulation tool for particle-based materials
  modeling at the atomic, meso, and continuum scales.
\newblock \emph{Computer Physics Communications}, 271:\penalty0 108171, 2022.

\bibitem[Stukowski(2010)]{Stukowski2009}
A.~Stukowski.
\newblock Visualization and analysis of atomistic simulation data with
  {OVITO}---the {Open Visualization Tool}.
\newblock \emph{Modelling and Simulation in Materials Science and Engineering},
  18:\penalty0 015012, 2010.

\bibitem[Ronneberger et~al.(2015)Ronneberger, Fischer, and
  Brox]{Ronneberger2015}
O.~Ronneberger, P.~Fischer, and T.~Brox.
\newblock {U-Net}: Convolutional networks for biomedical image segmentation.
\newblock In \emph{International Conference on Medical Image Computing and
  Computer-Assisted Intervention}, pages 234--241. Springer, 2015.

\bibitem[Vaswani et~al.(2017)]{Vaswani2017}
A.~Vaswani et~al.
\newblock Attention is all you need.
\newblock \emph{Advances in Neural Information Processing Systems},
  30:\penalty0 5998--6008, 2017.

\bibitem[Rombach et~al.(2022)Rombach, Blattmann, Lorenz, Esser, and
  Ommer]{Rombach2022}
R.~Rombach, A.~Blattmann, D.~Lorenz, P.~Esser, and B.~Ommer.
\newblock High-resolution image synthesis with latent diffusion models.
\newblock In \emph{Proceedings of the IEEE/CVF Conference on Computer Vision
  and Pattern Recognition}, pages 10684--10695, 2022.

\end{thebibliography}


\begin{thebibliography}{9}

\bibitem{Ho2022CFG}
J. Ho and T. Salimans,
``Classifier-free diffusion guidance,''
\textit{arXiv preprint arXiv:2207.12598} (2022).

\bibitem{Liu2014}
K. Liu and A. Tovar,
``An efficient 3D topology optimization code written in Matlab,''
\textit{Structural and Multidisciplinary Optimization} \textbf{50},
1175--1196 (2014).

\bibitem{Mishin1999}
Y. Mishin, D. Farkas, M. Mehl, and D. Papaconstantopoulos,
``Interatomic potentials for monoatomic metals from experimental data
and ab initio calculations,''
\textit{Physical Review B} \textbf{59}, 3393 (1999).

\end{thebibliography}

% ...

\section*{Acknowledgements}

This work used Expanse at SDSC through allocation MAT230081 from the
Advanced Cyberinfrastructure Coordination Ecosystem: Services \& Support
(ACCESS) program, which is supported by U.S. National Science Foundation
grants \#2138259, \#2138286, \#2138307, \#2137603, and \#2138296.

\section*{Author contributions}

C.-T.C.\ conceived the idea, designed the theory and modeling approach,
and implemented the simulations. C.-T.C.\ and D.L.\ analyzed and
interpreted the data. C.-T.C.\ wrote the original draft, and D.L.\
revised and edited the manuscript.

\section*{Competing interests}

The authors declare no competing interests.

\clearpage
\newcommand{\SIincluded}{}
\ifdefined\SIincluded\else

\documentclass[12pt]{article}

\usepackage[margin=1in]{geometry}
\usepackage{amsmath,amssymb,amsthm}
\usepackage{mathtools}
\usepackage{bm}
\usepackage{graphicx}
\usepackage{xcolor}
\usepackage{hyperref}
\usepackage[numbers]{natbib}
\usepackage{setspace}
\usepackage{microtype}
\usepackage{booktabs}
\usepackage{caption}
\usepackage{subcaption}
\usepackage{float}
\usepackage{array}

\newcommand{\eps}{\varepsilon}
\newcommand{\Ebar}[1]{\overline{\widetilde{E}}_{#1}}
\newcommand{\Shat}{\widehat{S}}
\newcommand{\Hhat}{\widehat{H}}
\newcommand{\rmin}{r_{\min}}
\newcommand{\dist}{\mathrm{dist}}

\doublespacing

\title{\textbf{Supplementary Information}\\[6pt]
\large Scaling atom-by-atom inverse design with nano-topology optimization
and diffusion models}

\author{Chun-Teh Chen$^{1}$ \and Denvid Lau$^{2}$}

\date{
  \small $^{1}$Department of Materials Science and Engineering,
  University of California, Berkeley, CA, USA\\
  \small $^{2}$Department of Architecture and Civil Engineering,
  City University of Hong Kong, Hong Kong, China}

\begin{document}

\maketitle

\fi 

\providecommand{\kbulk}{k_{\mathrm{bulk}}}
\providecommand{\kbulkstar}{k_{\mathrm{bulk}}^{*}}

\renewcommand{\theequation}{S\arabic{equation}}
\renewcommand{\thefigure}{S\arabic{figure}}
\renewcommand{\thetable}{S\arabic{table}}
\renewcommand{\thepage}{S\arabic{page}}
\setcounter{page}{1}
\setcounter{figure}{0}
\setcounter{table}{0}
\setcounter{equation}{0}

\renewcommand{\theHequation}{SI.\arabic{equation}}
\renewcommand{\theHfigure}{SI.\arabic{figure}}
\renewcommand{\theHtable}{SI.\arabic{table}}
\renewcommand{\theHsection}{SI.\arabic{section}}

\renewcommand{\thesection}{A.\arabic{section}}
\setcounter{section}{0}

\noindent\textbf{This Supplementary Information contains:}

\medskip
\noindent\textbf{A\quad Supplementary Notes}

\begin{quote}
  \textbf{A.1}\quad Surface elasticity of nanowires\\
  \textbf{A.2}\quad Crystallography-aligned multi-shell sensitivity filter\\
  \textbf{A.3}\quad Stability of using a local sensitivity filter\\
  \textbf{A.4}\quad Mirror symmetry in FCC crystals\\
  \textbf{A.5}\quad Reference designs for nanocantilevers\\
  \textbf{A.6}\quad Design of nanocantilevers without mirror symmetry\\
  \textbf{A.7}\quad Choosing classifier-free guidance strength for Gaussian-DDPM\\
  \textbf{A.8}\quad Choosing classifier-free guidance strength for TO-DDPM\\
  \textbf{A.9}\quad Similarity between generated designs and Nano-TO training samples\\
  \textbf{A.10}\quad Multi-objective selection using TO-DDPM\\
  \textbf{A.11}\quad Design of finite-thickness nanocantilevers\\
  \textbf{A.12}\quad Bulk-equivalent bending stiffness reference\\
  \textbf{A.13}\quad Design of finite-thickness nanocantilevers at reduced scale\\
  \textbf{A.14}\quad Nanopillars designed by FEM-TO versus Nano-TO%
  \ifdefined\SIincluded\else\\
  \textbf{A.15}\quad Supplementary references%
  \fi
\end{quote}

\clearpage

\section*{A\quad Supplementary Notes}

\subsection*{A.1\quad Surface elasticity of nanowires}

We investigate aluminum nanowires with axial orientations $[100]$,
$[110]$, and $[111]$, whose effective Young's moduli deviate from bulk
values as the wire radius decreases. The size dependence seen in
Figure~S1 arises from surface elasticity: free surfaces behave as
two-dimensional elastic media that modify the effective response of a
finite body. In the thin-body regime ($t \ll L$, $t \gg a_0$, where $t$
is the thickness of a slab, $L$ is a characteristic in-plane dimension,
and $a_0$ is an interatomic spacing), the in-plane effective Young's
modulus of a slab for uniaxial loading along a unit vector $\mathbf{d}$
lying in the surface, denoted $E_{(hkl)}(\mathbf{d};t)$, with
$\mathbf{n}\cdot\mathbf{d} = 0$, admits the leading-order expansion:

\begin{equation}
  E_{(hkl)}\!\left(\mathbf{d};t\right)
  = E_{\mathrm{bulk}}\!\left(\mathbf{d}\right)
    + \frac{2}{t}K_s^{(hkl)}\!\left(\mathbf{d}\right)
    + O\!\left(t^{-2}\right)
  \label{eq:SI_Ehkl}
  \tag{S1}
\end{equation}

where $K_s^{(hkl)}$ is the surface tangent stiffness (N\,m$^{-1}$). A
negative slope of $E_{(hkl)}$ versus $t$ indicates a surface stronger
than the bulk along $\mathbf{d}$; a positive slope indicates a weaker
surface. The effective Young's moduli of nano-slabs in Figure~S2 show
direction-resolved surface signatures that determine the sign and
magnitude of surface elasticity.

Another essential element is symmetry. Each low-index surface of a cubic
crystal is itself a two-dimensional crystal with a specific point group.
That symmetry constrains the in-plane elastic tensor and, crucially,
enforces degeneracies between directions related by symmetry operations.
The three surfaces relevant here, \{100\}, \{110\}, and \{111\}, exhibit
distinct in-plane symmetry: $C_{4v}$ for \{100\} (square), $C_{2v}$ for
\{110\} (rectangular), and $C_{3v}$ for \{111\} (triangular). Directions
that are symmetry-equivalent must have identical moduli at a given
thickness; directions that are not symmetry-related are free to differ.
This symmetry principle explains which curves coincide within each panel
of Figure~S2 and which do not.

The \{100\} plane has four-fold rotations and mirrors (point group
$C_{4v}$). The axial pair
$[010]/[001]$ is symmetry-equivalent (hence coincident curves), and the
diagonal pair $[01\bar{1}]/[011]$ is also symmetry-equivalent. However,
no symmetry maps an axis to a diagonal; therefore, the two families may
differ. In our case, the axial family increases with thickness (surfaces
weaker than the bulk), whereas the diagonal family decreases (surfaces
stronger than the bulk), resulting in two distinct plateaus in
Figure~S2a. The equality within each pair and the inequality between the
two pairs follow directly from $C_{4v}$ symmetry.

The \{110\} plane has two-fold rotations and mirrors (point group
$C_{2v}$). The orthogonal pair $[1\bar{1}0]/[001]$ is not
symmetry-equivalent and can have different moduli and different slopes
with thickness. In our case, the $[1\bar{1}0]$ curve decreases with $t$
(surfaces stronger than the bulk), whereas the $[001]$ curve increases
with $t$ (surfaces weaker than the bulk). The rotated orthogonal pair
$[1\bar{1}2]/[\bar{1}11]$, although sharing a common decreasing trend
with $t$ (surfaces stronger than the bulk), has different moduli and
different slopes with thickness (Figure~S2b).

For cubic crystals, restricting $E(\mathbf{n})$ to the \{111\} plane
makes it independent of the in-plane angle: the cubic invariant
$\beta = n_1^2 n_2^2 + n_2^2 n_3^2 + n_3^2 n_1^2$ becomes
$\beta = (n_1 n_2 + n_2 n_3 + n_3 n_1)^2 = \tfrac{1}{4}$ when
$n_1 + n_2 + n_3 = 0$. Therefore, any orthogonal pair chosen within \{111\} is
degenerate (with the same in-plane modulus). For instance, the orthogonal pair
$[1\bar{1}0]/[11\bar{2}]$ and the rotated pair
$[1\bar{2}1]/[10\bar{1}]$ have the same curves (Figure~S2c). Both
collapse to a single curve at each thickness and exhibit the same
decreasing trend with $t$ (surfaces stronger than the bulk).

A nanowire loaded along axis $\mathbf{e}$ samples the surface elasticity
of each lateral facet through a geometric projection onto $\mathbf{e}$
and a perimeter-to-area scaling. For a circular cross-section:

\begin{equation}
  E_{\mathrm{wire}}\!\left(\mathbf{e};r\right)
  \approx E_{\mathrm{bulk}}\!\left(\mathbf{e}\right)
    + \frac{P}{A}
      \bigl\langle k_s(\mathbf{e})\bigr\rangle_{\mathrm{facets}}
  \label{eq:SI_Ewire}
  \tag{S2}
\end{equation}

with $P/A = 2/r$; $\langle\cdot\rangle$ denotes a perimeter-average
weighted by facet length. The sign of the size effect (whether smaller is stronger or weaker) is therefore set
by the net surface contrast accumulated over all bounding surfaces,
evaluated in the axial loading direction.

For $[100]$ wires, the lateral facets are \{100\} and \{110\}, as
shown in Figure~S1d. The \{100\} contribution from the axial family
(weaker-than-bulk) (Figure~S2a, left), together with the \{110\}
response (weaker-than-bulk) (Figure~S2b, left), results in a negative
perimeter-averaged surface contrast along $[100]$, producing the
``smaller-is-weaker'' trend in Figure~S1a. On \{100\}, symmetry makes
the axial pair $[010]/[001]$ and the diagonal pair $[01\bar{1}]/[011]$
each degenerate (identical curves within a pair). On \{110\}, the
directions $[1\bar{1}0]$ and $[001]$ are not symmetry-equivalent, so
mixed trends are allowed.

For $[110]$ wires, the lateral facets are \{100\}, \{110\}, and
\{111\}, as shown in Figure~S1e. The \{100\} diagonal family relevant
to $[110]$ is stronger-than-bulk (Figure~S2a, right); the \{110\}
facets contribute a positive surface contrast when projected onto the
$[110]$ axis (Figure~S2b, left); and the isotropic \{111\} facets
contribute a uniformly positive correction (Figure~S2c). The resulting
perimeter-averaged surface contrast is positive, yielding the
``smaller-is-stronger'' trend in Figure~S1b. Symmetry explains why the
\{111\} facet's contribution is azimuth-independent.

For $[111]$ wires, the lateral facets are \{110\} alone, as shown in
Figure~S1f. On \{110\}, the in-plane direction parallel to
$[\bar{1}11]$ shows a stronger-than-bulk surface response
(Figure~S2b, right). Therefore, the perimeter-averaged surface
contrast is positive, giving the ``smaller-is-stronger'' trend in
Figure~S1c.

\begin{figure}[H]
  \centering
  \includegraphics[width=0.92\textwidth]{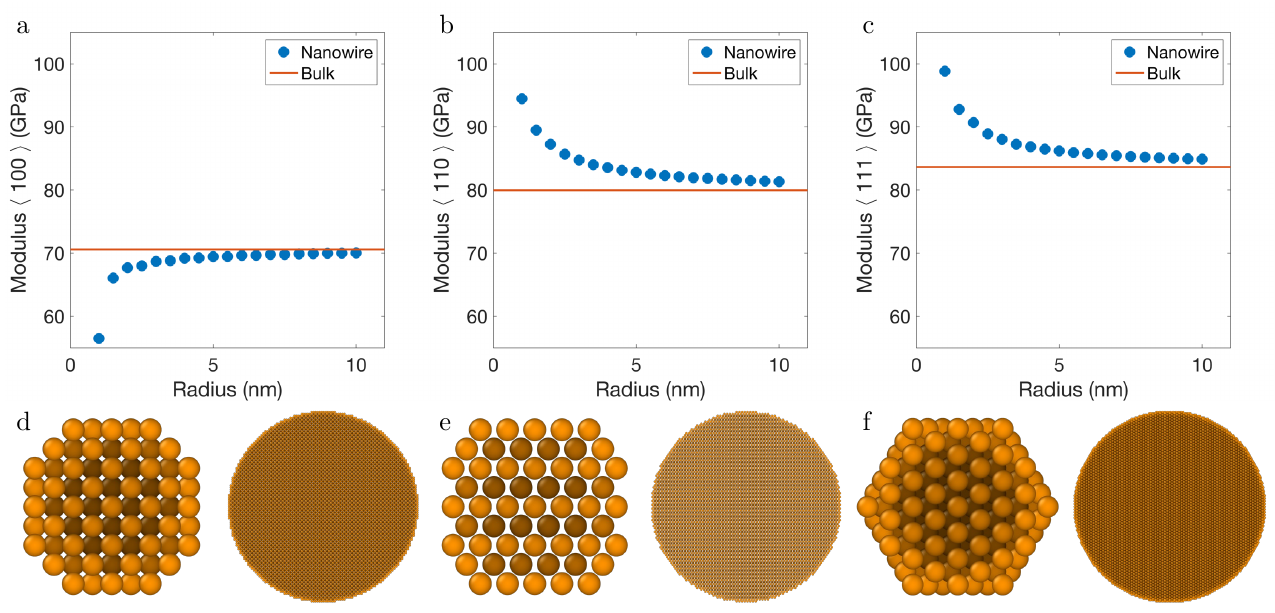}
  \caption{\textbf{Surface effects on the elasticity of nanowires.}
    \textbf{a}, Nanowires oriented along $[100]$ become softer as the
    radius decreases, resulting in a ``smaller is weaker'' effect.
    \textbf{b} and \textbf{c}, Nanowires oriented along $[110]$ and
    $[111]$ become stiffer as the radius decreases, resulting in a
    ``smaller is stronger'' effect. \textbf{d}--\textbf{f}, Cross
    sections of nanowires oriented along $[100]$, $[110]$, and $[111]$,
    respectively. The smallest (radius $\approx 1$~nm) and largest
    (radius $\approx 10$~nm) cross sections are shown.}
  \label{fig:S1}
\end{figure}

\begin{figure}[H]
  \centering
  \includegraphics[width=0.62\textwidth]{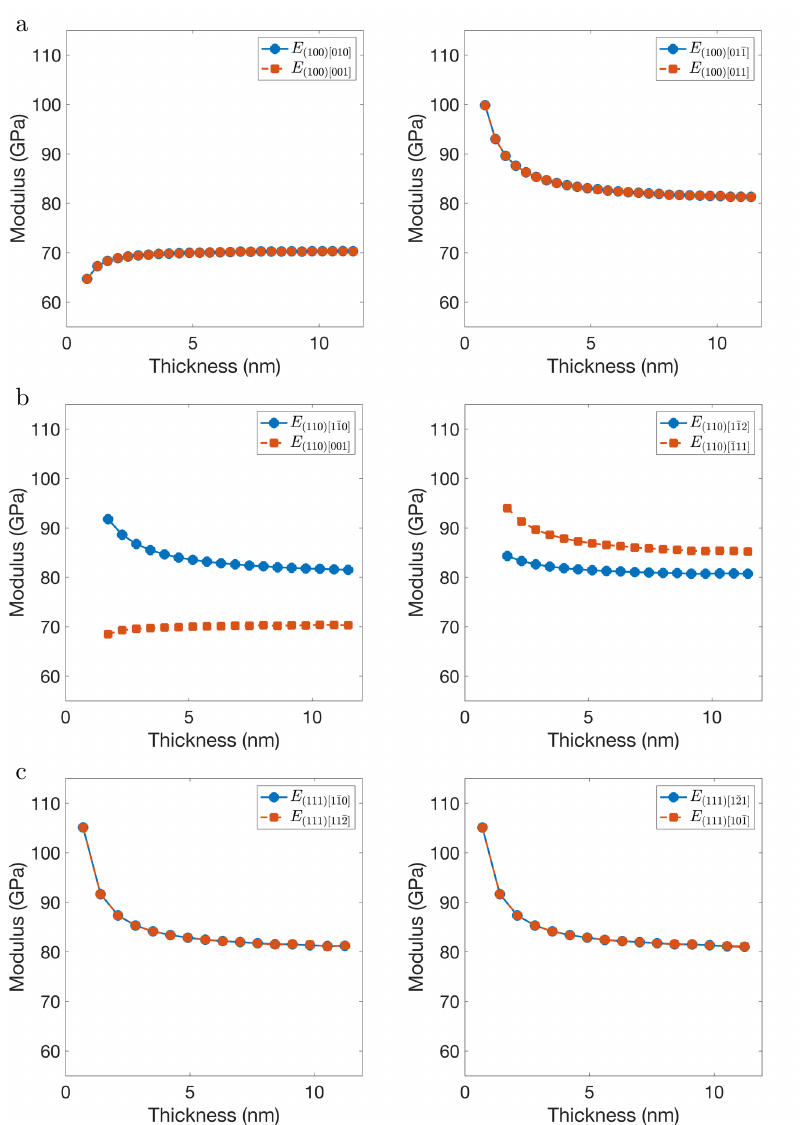}
  \caption{\textbf{Surface effects on the elasticity of nano-slabs.}
    \textbf{a}--\textbf{c}, Effective Young's moduli of nano-slabs
    with \{100\}, \{110\}, and \{111\} surfaces in two pairs of
    in-plane directions, respectively. For \{100\}, \{110\}, and \{111\}
    nano-slabs, the right panels show rotated orthogonal in-plane pairs
    relative to the left panels, by 45$^\circ$, $\approx 35.3^\circ$,
    and 60$^\circ$, respectively (directions labeled in the legends).}
  \label{fig:S2}
\end{figure}

\subsection*{A.2\quad Crystallography-aligned multi-shell sensitivity filter}

The filter covers the first twelve face-centered cubic (FCC) neighbor
shells with a total of 248 atoms. The filtered sensitivity value of atom
$k$ is calculated as:

\begin{equation}
  \Shat_k
  = \frac{1}{\displaystyle\sum_{i=1}^{n}\Hhat_i}
    \sum_{i=1}^{n}
    \Hhat_i \frac{\Ebar{i}}{x_i}
  \tag{S3a}
\end{equation}

\begin{equation}
  \Hhat_i
  = \rmin - \dist(k,i),
  \quad
  \bigl\{i \in \mathcal{N}(k) \mid \dist(k,i) \le \rmin\bigr\}
  \tag{S3b}
\end{equation}

where $\mathcal{N}(k)$ is the fixed neighborhood of atom $k$, $\Hhat_i$
is the weighting factor for atom $i$, $\dist(k,i)$ is the distance
between atoms $k$ and $i$, and $\rmin$ is the filter radius. Table~S1
provides the shell radii and multiplicities. The filter radius is set to
10.325~\AA, the same as the radius of the 13th-neighbor shell.
Therefore, the 13th-neighbor shell is not considered in the calculation
as its weighting factor is zero ($\rmin = \dist$). Figure~S3a shows the
positions of neighbor atoms around a reference atom in each shell. To
illustrate the filter's effect, we create a cube with 32{,}000 atoms
and calculate its sensitivity map. Figure~S3b shows the cube and an
$x$--$z$ slice, with real atoms in orange and virtual atoms in blue. We
calculate per-atom sensitivities to $z$-directional stiffness. The raw
sensitivity map (Figure~S3c, left) shows a sharp jump at the boundaries
between real and virtual atoms, as all virtual atoms have zero raw
sensitivity values. After applying the crystallography-aligned
multi-shell sensitivity filter, the filtered sensitivity map
(Figure~S3c, right) exhibits a smooth transition across the boundaries:
virtual atoms gain nonzero sensitivity values when real atoms are within
their 12th-neighbor shells, and boundary artifacts are suppressed. This
filtered sensitivity map produces a more stable signal for subsequent
optimization.

\begin{table}[H]
  \centering
  \caption{\textbf{Radii and multiplicities of the first twelve neighbor
    shells in an aluminum lattice.} Values are measured from a reference
    atom (counts exclude the reference atom).}
  \label{tab:S1}
  \begin{tabular}{lcccccccccccc}
    \toprule
    Shell      & 1 & 2 & 3 & 4 & 5 & 6 & 7 & 8 & 9 & 10 & 11 & 12 \\
    \midrule
    $r$ (\AA)  & 2.864 & 4.050 & 4.960 & 5.728 & 6.404 & 7.015
               & 7.577 & 8.100 & 8.591 & 9.056 & 9.498 & 9.920 \\
    Count      & 12 & 6 & 24 & 12 & 24 & 8 & 48 & 6 & 36 & 24 & 24 & 24 \\
    \bottomrule
  \end{tabular}
\end{table}

\begin{figure}[H]
  \centering
  \includegraphics[width=0.92\textwidth]{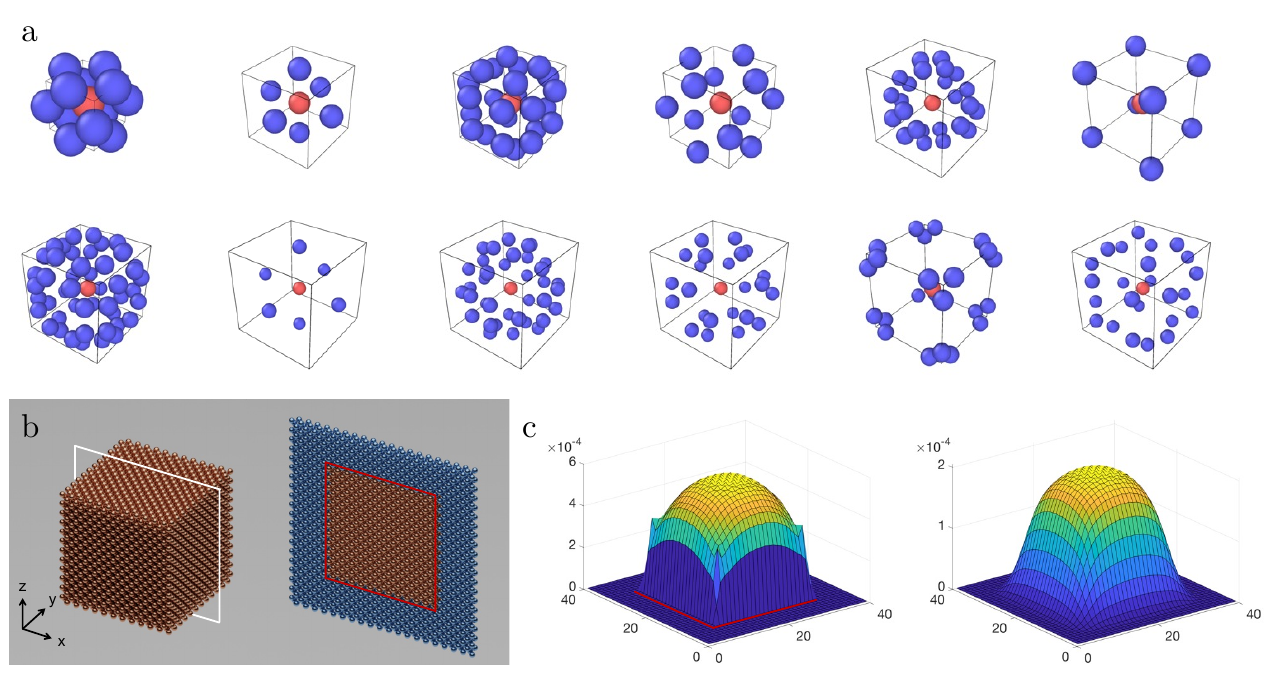}
  \caption{\textbf{Physics-aligned nonlocal sensitivity filter and its
    effects on the sensitivity map.}
    \textbf{a}, First twelve FCC neighbor shells around a reference atom
    (red). These twelve shells lie within the filter radius of 10.325~\AA\
    (see Table~S1 for shell radii and counts).
    \textbf{b}, An aluminum cube with 32{,}000 atoms. Left: model with
    the $x$--$z$ slicing plane (white frame). Right: $x$--$z$ slice;
    real atoms are in orange and virtual atoms (padding) are in blue; red
    outline marks the boundaries between real and virtual atoms.
    \textbf{c}, Sensitivity map of the slice before filtering (left) and
    after applying the filter (right). The raw sensitivity map shows a
    sharp jump on the boundaries due to zero-valued virtual atoms. In
    contrast, the filtered sensitivity map exhibits a smooth transition
    and non-zero values near the boundaries, leading to a more reliable
    signal for subsequent optimization.}
  \label{fig:S3}
\end{figure}

\subsection*{A.3\quad Stability of using a local sensitivity filter}

The ``local'' filter averages sensitivities only over the first FCC
neighbor shell. The filter radius is set to 4.05~\AA, the same as the
radius of the second-neighbor shell. Therefore, the second-neighbor
shell is not considered in the calculation as its weighting factor is
zero ($\rmin = \dist$). We apply this filter to perform the
same design task for nanocantilevers under thickness-periodic boundary
conditions (PBCs) as described in the main text. Figure~S4 shows the
terminal states from eight independent trials just before failure
(``lost atoms'' in LAMMPS). None of the trials reaches the target mass
ratio of 59.60\%; they stall between 83.68\% and 65.47\%. A common
failure mode is the formation of a percolating void network, leaving
disconnected islands of real atoms (``floating'' atoms) that no longer
carry load.

In the embedded-atom method (EAM), energy depends on neighbors across
multiple shells. Restricting the filter to the first shell produces a
high-variance, speckled sensitivity map. When many atoms are flipped per
iteration (large-batch updates), fine-scale fluctuations translate into
scattered removals, which quickly connect into void channels and break
connectivity. Using a larger, multi-shell filter (e.g., 10.325~\AA\ in
the main text, covering the first 12 shells) averages sensitivities over
the physically relevant neighborhood, damping atom-scale noise and
leading to spatially coherent updates. A filter also sets an effective
minimum feature size. A larger filter reduces the formation of isolated
real-atom islands and prevents premature void percolation, enabling the
structure to shed mass while keeping load-bearing paths. Additionally,
first-order sensitivity updates are more reliable when the update
direction is smooth. Smoothing the sensitivity functions acts as a low-pass
filter. Therefore, large-batch updates stay aligned with the underlying
objective rather than reacting to local fluctuations.

\begin{figure}[H]
  \centering
  \includegraphics[width=0.92\textwidth]{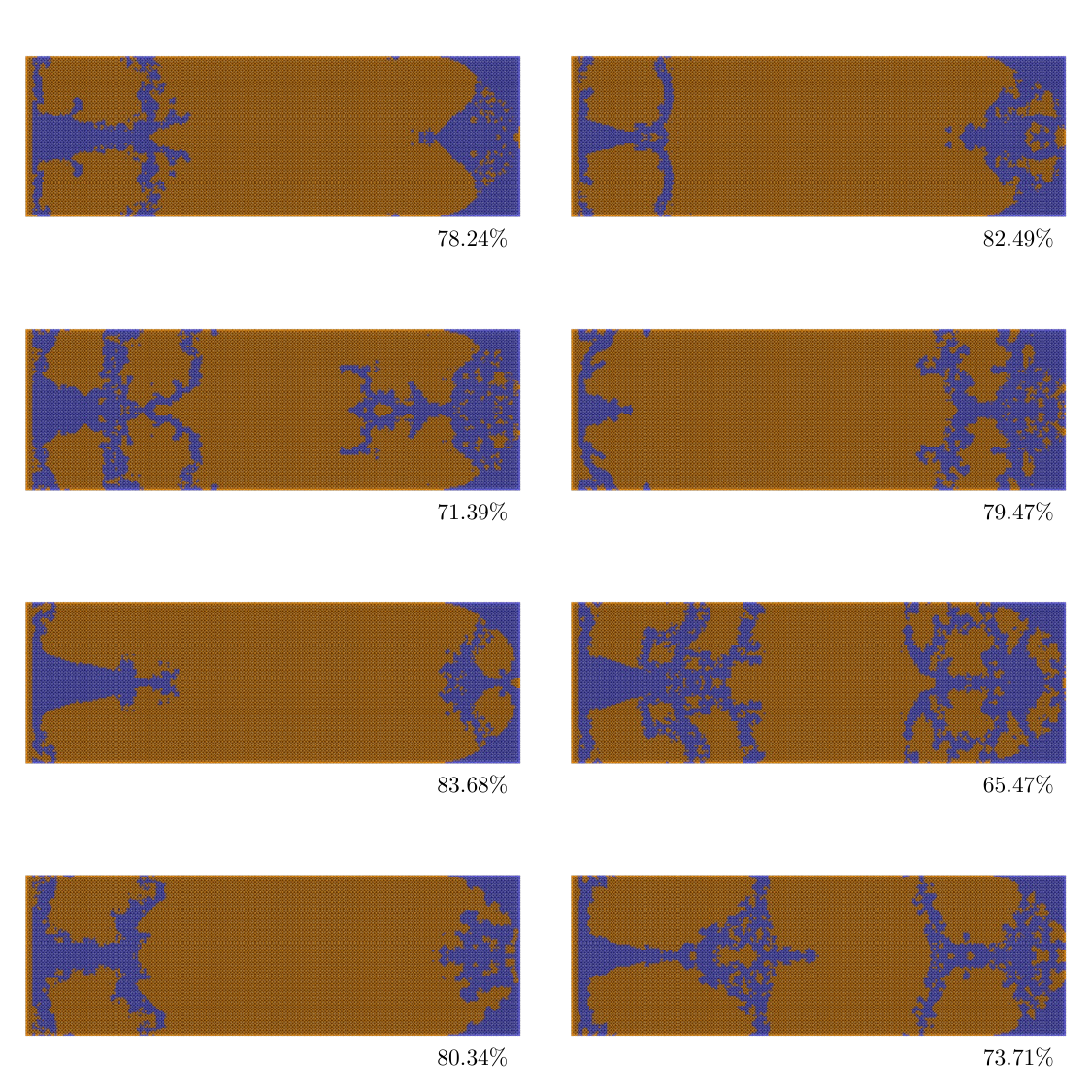}
  \caption{\textbf{Optimization instability caused by a local
    sensitivity filter.} Final configurations from eight independent
    trials for nanocantilevers under thickness-periodic boundary
    conditions using a local sensitivity filter. Real atoms are in
    orange, and virtual atoms are in blue. Numbers at the bottom of each
    configuration represent the achieved mass ratios; none reach the
    59.60\% target. In each trial, a percolating void network forms,
    leaving disconnected islands of real atoms and resulting in ``lost
    atoms'' termination in LAMMPS.}
  \label{fig:S4}
\end{figure}

\subsection*{A.4\quad Mirror symmetry in FCC crystals}

In FCC crystals viewed along $[100]$, the crystal is built from atomic
layers that alternate laterally: an A layer is followed by a B layer.
The A layer occurs at $x = na$, where $a$ is the lattice constant.
At $x = 0$, the in-plane $y$--$z$ motif repeats every $a$ in $y$ and
$z$. The B layer occurs at $x = na + a/2$. At $x = a/2$, the in-plane
$y$--$z$ motif again repeats every $a$ in $y$ and $z$. The B layer is
the A layer shifted in the $y$--$z$ plane by $a/2$ along $y$ or $z$. Since the A
layer and the B layer differ by this lateral translation, a slab with
the bottom surface as A and the top surface as B is not mirror-symmetric
about the mid-plane: a reflection through the mid-plane flips
$x\to -x$ but leaves $y,z$ unchanged. Thus, atoms on the bottom surface
do not map onto atoms on the top surface without an additional in-plane
shift.

We can make the slab nearly mirror-symmetric by trimming one terminal
half-cell so that both exposed surfaces end on the same registry. As
explained in Figure~S5, removing the top B layer leaves A terminations
on both sides. Then, the two halves of the slab are related by a
mid-plane reflection. Therefore, surfaces on the top half are identical
to those on the bottom half. In discussing that symmetry, atoms lying
exactly on the mid-plane are self-mapped by the mirror and thus cannot
by themselves establish the equivalence of the two surfaces. Here, we
treat such mid-plane atoms as self-pairs (e.g., atoms 7 and 8 in
Figure~S5), ensuring every atom in the slab participates in the
symmetry mapping.

\begin{figure}[H]
  \centering
  \includegraphics[width=0.92\textwidth]{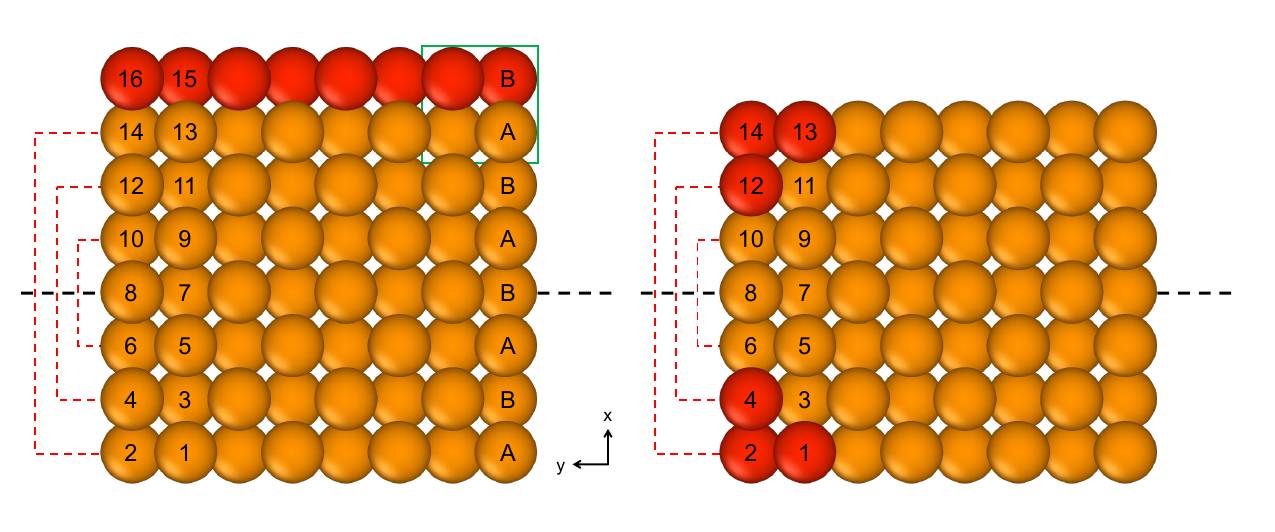}
  \caption{\textbf{Constructing a mirror-symmetric nano-slab by trimming
    the terminal half-cell.} Left: A schematic of an FCC nano-slab with
    A/B stacking (labels on the right). The top B layer is highlighted in
    red; keeping it results in A/B terminations that are not related by a
    mirror reflection. The green box marks a unit cell containing A/B
    layers. Right: After removing the top B layer, both exposed surfaces
    end on the same registry, giving A/A terminations. The nano-slab is
    now invariant under reflection in the mid-plane (black dashed line):
    atoms above and below the plane form mirror pairs (red dashed
    brackets). Atoms that lie on the mid-plane (e.g., atoms 7 and 8)
    are self-mapped by the mirror and are treated as self-pairs in the
    symmetry mapping.}
  \label{fig:S5}
\end{figure}

\subsection*{A.5\quad Reference designs for nanocantilevers}

To evaluate the performance of Nano-TO designs, we construct reference
designs by uniformly reducing the beam height while keeping the length
and width fixed. The mass ratio equals the height ratio $h/h_0$. For
thickness-periodic nanocantilevers, the reference design domain is
200.475$\times$20.25$\times$615.60~\AA\ with 150{,}480 atoms:
148{,}500 active atoms and 1{,}980 passive atoms at the clamped
boundary. By converting active atoms from real to virtual based on their
height coordinates, we create beams with target mass ratios.
Figure~S6 shows the height-scaled reference beams with mass ratios of
75.76\%, 67.68\%, and 59.60\%, corresponding to converting 36{,}000,
48{,}000, and 60{,}000 real atoms into virtual atoms, respectively.
The corresponding heights are 151.875, 135.675, and 119.475~\AA,
respectively ($h = \mu h_0$ with $h_0 = 200.475$~\AA).

For small-deflection bending of a cantilever, the Euler--Bernoulli
theory gives:

\begin{equation}
  k = \frac{3EI}{L^3}
  \tag{S4a}
\end{equation}

For a rectangular section, the second moment of area (area moment of inertia) is:

\begin{equation}
  I = \frac{bh^3}{12}
  \tag{S4b}
\end{equation}

Thus, with material $E$, length $L$, and width $b$ fixed, the bending
stiffness scales as $k \propto h^3$. Relative to the initial beam, the
estimated bending stiffness of a reference design is:

\begin{equation}
  \frac{k}{k_0}
  = \left(\frac{h}{h_0}\right)^3
  = \mu^3
  \tag{S4c}
\end{equation}

This relation is used to estimate the stiffness of the height-scaled
reference designs.

\begin{figure}[H]
  \centering
  \includegraphics[width=0.92\textwidth]{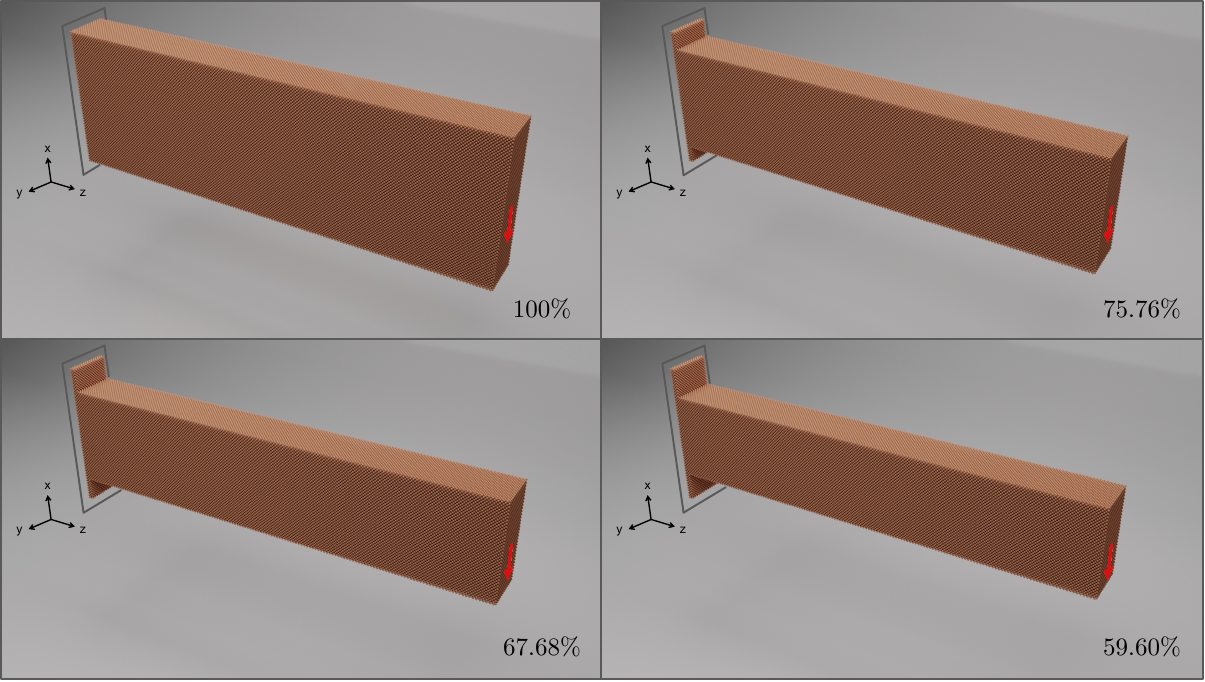}
  \caption{\textbf{Height-scaled reference beams.} Initial design
    (100\%) and three height-scaled reference beams at different mass
    ratios (75.76\%, 67.68\%, and 59.60\%). The corresponding heights
    are 200.475, 151.875, 135.675, and 119.475~\AA, respectively. The
    gray block represents the clamped support; the red arrow marks the
    applied vertical displacement at the free end. Models are rendered
    with three periodic images in the thickness direction.}
  \label{fig:S6}
\end{figure}

\subsection*{A.6\quad Design of nanocantilevers without mirror symmetry}

To assess the role of symmetry, we repeat the nanocantilever design task
without enforcing mirror symmetry. All other settings (thickness-periodic
boundary, two-phase update schedule, filtering) are identical to those in
the main text, and we run 64 independent trials. Figure~S7 shows the
initial design (100\%) and the best designs at mass ratios of 75.76\%,
67.68\%, and 59.60\%. The optimized layouts feature asymmetric truss-like
motifs. Figure~S8 reports the normalized bending stiffness for all 64
trials at each mass ratio (colored circles), relative to the initial
design. As a baseline, we compare our results against the
height-scaled reference designs, whose stiffness is obtained from
atomistic simulations (blue dots) and estimated using the
Euler--Bernoulli theory (red curve), as described in A.5. The
optimized, asymmetric designs consistently
exceed the reference designs at the same mass ratio. At a mass ratio of
59.60\%, the best normalized stiffness is 0.818, slightly below 0.820
achieved with mirror symmetry (main text). Since allowing asymmetry
enlarges the design space, the unconstrained global optimum cannot be
worse than the symmetric one. The small shortfall reflects search
efficiency under a fixed compute budget. Imposing symmetry reduces the
number of design variables and avoids exploring left-right variants of
the same layout, which helps the optimizer reach a high-quality solution
more reliably.

\begin{figure}[H]
  \centering
  \includegraphics[width=0.92\textwidth]{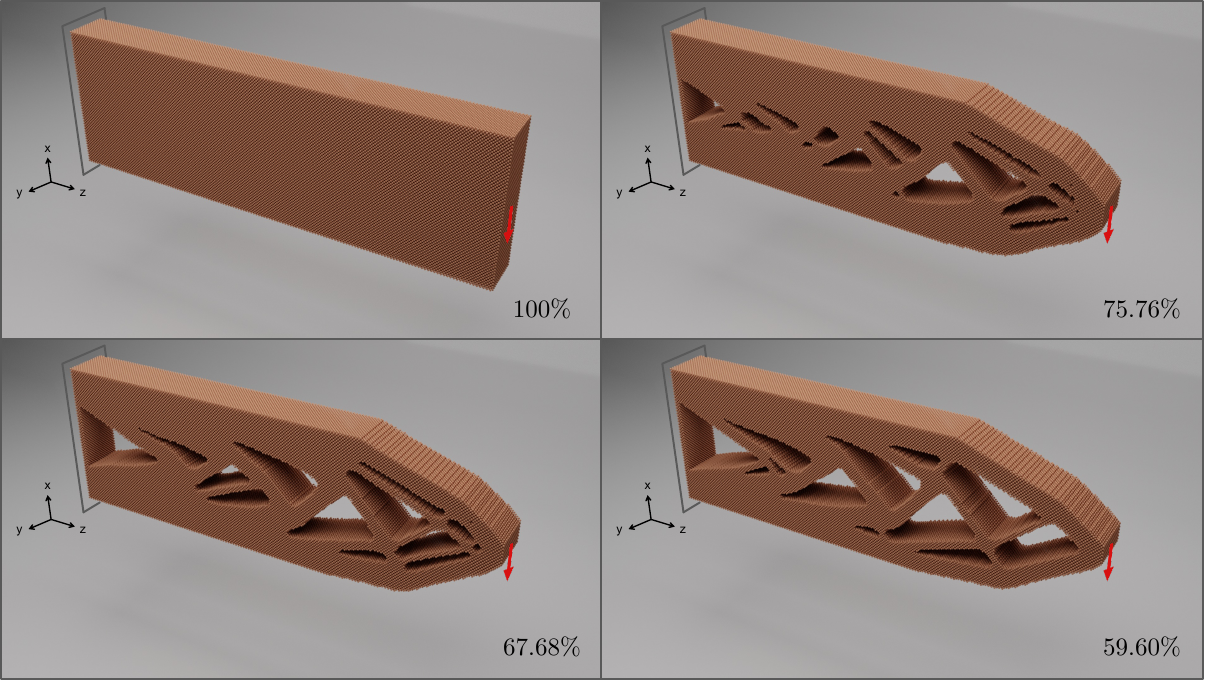}
  \caption{\textbf{Nano-TO design of thickness-periodic nanocantilevers
    without mirror symmetry.} Initial design (100\%) and optimized
    designs at different mass ratios (75.76\%, 67.68\%, and 59.60\%).
    The gray block represents the clamped support; the red arrow marks
    the applied vertical displacement at the free end. Models are
    rendered with three periodic images in the thickness direction.}
  \label{fig:S7}
\end{figure}

\begin{figure}[H]
  \centering
  \includegraphics[width=0.62\textwidth]{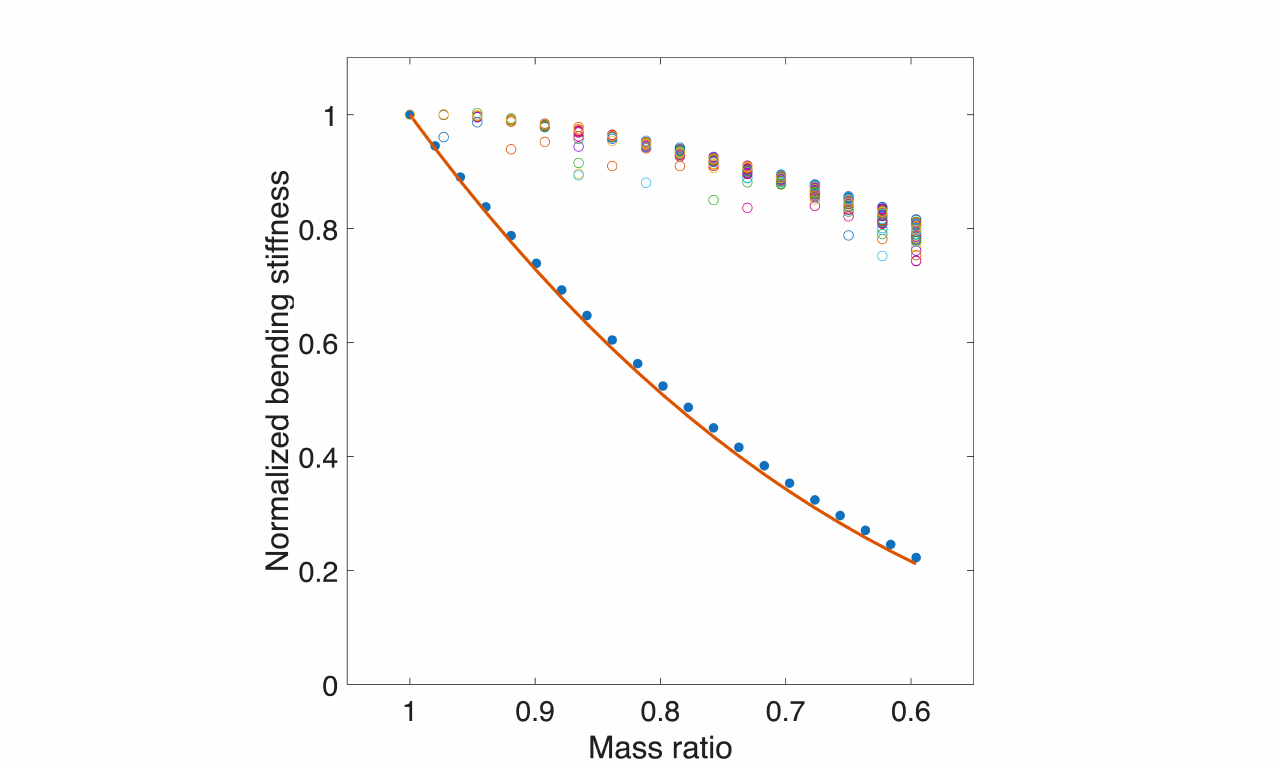}
  \caption{\textbf{Normalized bending stiffness of nanocantilevers
    without mirror symmetry versus mass ratio.} Colored circles: results
    from 64 independent trials at each mass ratio. Blue dots:
    height-scaled reference designs. Red curve: Euler--Bernoulli
    estimate.}
  \label{fig:S8}
\end{figure}

\subsection*{A.7\quad Choosing classifier-free guidance strength for Gaussian-DDPM}

We investigate the effects of the classifier-free guidance (CFG)
strength $w \in \{1, 3, 5, 7\}$ on inference, while fixing the
stiffness condition to $k = 1.0$. For each $w$, we generate 1{,}600
designs, convert each into an atomistic model with the same
thickness-periodic setup, and evaluate their normalized bending
stiffness. We report two metrics per $w$: the mean stiffness across the
generated set and the top 1\% stiffness. As shown in Figure~S9,
increasing $w$ from 1 to 3 significantly improves both metrics: the
mean stiffness increases from 0.49 to 0.61, and the top stiffness
increases from 0.65 to 0.70. Beyond $w = 3$, gains saturate: the mean
stiffness rises only modestly from 0.61 to 0.65 by $w = 7$, and the
top stiffness plateaus around 0.71. A higher CFG is more mode-seeking,
concentrating samples near the conditional modes, and reducing
diversity~\cite{Ho2022CFG}. Our objective is to screen high-quality, yet
varied, designs. Therefore, retaining diversity is valuable. We adopt
$w = 3$ for the statistics reported in the main text, as it delivers a
high mean stiffness with a top stiffness comparable to stronger
guidance.

\begin{figure}[H]
  \centering
  \includegraphics[width=0.77\textwidth]{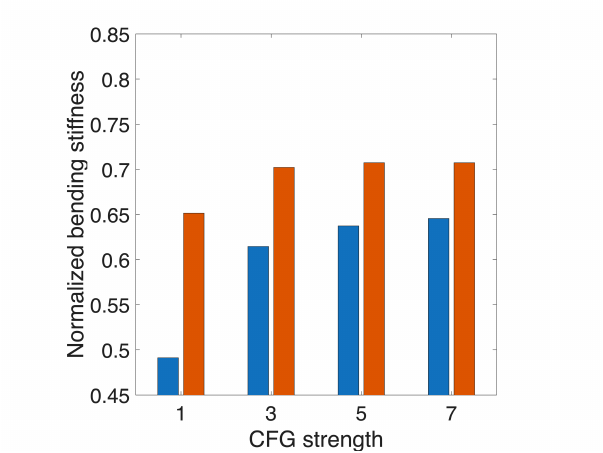}
  \caption{\textbf{Effects of classifier-free guidance for
    Gaussian-DDPM.} Mean (blue) and top (red) normalized bending
    stiffness of generated designs versus the CFG strength. The
    stiffness condition is set to 1.0.}
  \label{fig:S9}
\end{figure}

\subsection*{A.8\quad Choosing classifier-free guidance strength for TO-DDPM}

We investigate the effects of the stiffness condition $k \in \{0.0,
-0.2, -0.4, -0.6, -0.8, -1.0\}$ and CFG strength $w \in \{1, 3, 5,
7\}$ on inference, while fixing the mass-ratio condition to $m = -1.0$
(corresponding to a mass ratio of 59.60\%). For each $(k, w)$ pair, we
generate 1{,}600 designs, convert each into an atomistic model with the
same thickness-periodic setup, and evaluate their normalized bending
stiffness.

Unlike Gaussian-DDPM, Figure~S10 shows that increasing $w$ does not
increase the mean stiffness or the top stiffness. In TO-DDPM, the
training pool contains nearly optimal designs at the target mass ratio,
and stronger guidance makes sampling more focused on a narrow set of
geometries that the model already prefers. Thus, the mean stiffness does
not improve and can dip slightly. The top stiffness varies little with
$w$. At $m = -1.0$, the achievable upper envelope is physically
constrained and learned by the model. Adjusting $w$ mainly changes how
often we sample near that ceiling, not how high it is. In the
1{,}600-sample hyperparameter-selection sweep, $k = -0.6$ and $w = 1$
give one of the highest mean stiffness values (0.809) and the top 1\%
stiffness values (0.822), and are therefore chosen for the larger
32{,}000-sample production run reported in the main text. The maximum
stiffness of 0.860 reported in the main text is obtained from that
larger production run, not from the 1{,}600-sample selection sweep. By
our linear normalization, $k = -0.6$ corresponds to a target normalized
stiffness
of 0.823, which is close to the achievable upper envelope and helps
raise the mean stiffness. Using the smaller guidance strength $w = 1$
also preserves diversity, which is beneficial for downstream screening
under additional criteria (e.g., surface ratio, potential energy).

\begin{figure}[H]
  \centering
  \includegraphics[width=0.92\textwidth]{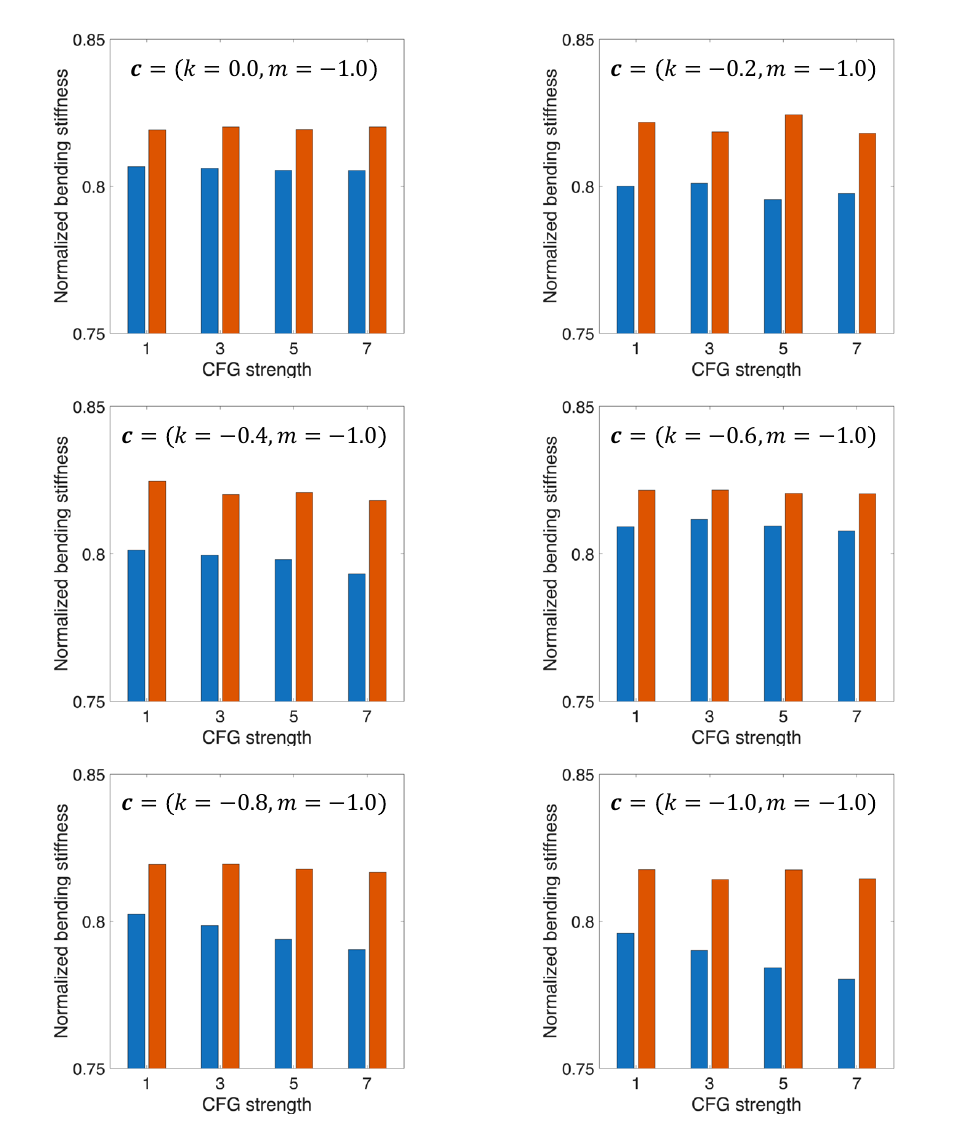}
  \caption{\textbf{Effects of classifier-free guidance for TO-DDPM.}
    Mean (blue) and top (red) normalized bending stiffness of generated
    designs versus the CFG strength and stiffness condition. The
    mass-ratio condition is set to $-1.0$.}
  \label{fig:S10}
\end{figure}

\subsection*{A.9\quad Similarity between generated designs and Nano-TO training samples}

We compare generated designs from TO-DDPM with Nano-TO training samples.
Since the 1{,}980 passive atoms at the clamp boundary do not change, we
only compare the 148{,}500 active atoms. Atom types are mapped to a
binary occupancy vector: virtual atoms are set to 0, and real atoms are
set to 1. This gives two binary matrices:
$X \in \{0,1\}^{m \times d}$ for training samples and
$Y \in \{0,1\}^{n \times d}$ for generated designs, where
$d = 148{,}500$. For each generated-training pair $(i,j)$, let
$A_{ij} = \sum_k Y_{ik} X_{jk}$ (positions where both have 1),
$B_{ij} = \sum_k Y_{ik}(1-X_{jk})$ (1 in $Y$, 0 in $X$),
$C_{ij} = \sum_k (1-Y_{ik}) X_{jk}$ (0 in $Y$, 1 in $X$), and
$D_{ij} = d - (A_{ij} + B_{ij} + C_{ij})$ (positions where both have
0). Percent identity (PID) is used to report similarity:
\begin{equation}
  \mathrm{PID}_{ij}
  = 100 \times
    \frac{A_{ij} + D_{ij}}{d}
  \tag{S5}
\end{equation}
Due to the mass ratio constraint of 59.60\%, the PID range is from
19.19\% to 100\%.

\subsection*{A.10\quad Multi-objective selection using TO-DDPM}

We select a high-performance TO-DDPM design (DM-07188) and compare it
against its nearest Nano-TO training samples. As shown in Figure~S11,
DM-07188 has nearest-neighbor similarities of 94.69\% and 94.02\% to
TO-44 and TO-15, respectively. DM-07188 has a normalized bending
stiffness of 0.822, which is 0.28\% above the best Nano-TO design and
1.30\% to 1.83\% above the nearest training samples. Compared to the
best Nano-TO design, both feature truss-like motifs with multiple
cross-braces. However, the best Nano-TO design features four
cross-braces, and DM-07188 has only three. This structural difference
results in a lower surface-atom fraction (0.1367 compared with 0.1436 for the
best Nano-TO design) and a potential energy per atom that is 0.0012~eV
lower than that of the best Nano-TO design, indicating improved energetic stability.

\begin{figure}[H]
  \centering
  \includegraphics[width=0.92\textwidth]{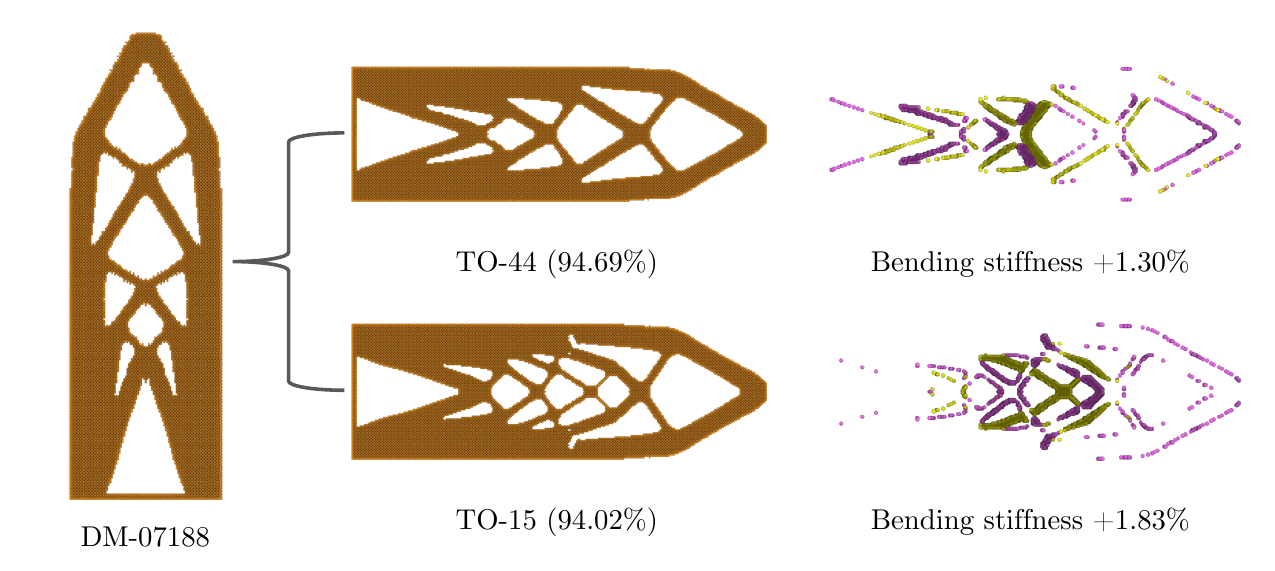}
  \caption{\textbf{Nearest-neighbor similarity and stiffness gain for a
    more stable nanocantilever.}
    DM-07188 is shown alongside its two nearest training samples,
    TO-44 (94.69\% PID) and TO-15 (94.02\% PID). DM-07188 achieves a normalized bending
    stiffness of 0.822, exceeding both nearest neighbors and the best Nano-TO design, while reducing
    the surface-atom fraction and lowering the potential energy per atom,
    indicating improved energetic stability.
    In each overlay, yellow marks atoms present only in the generated
    design and magenta marks atoms present only in the training sample;
    stiffness gains are relative to the respective training sample.}
  \label{fig:S11}
\end{figure}

\subsection*{A.11\quad Design of finite-thickness nanocantilevers}

In the main text, we impose periodic boundary conditions (PBCs) along
the thickness direction, removing side surfaces and approximating the
infinite-thickness limit. This choice aligns with the c-DDPM
cross-section representation and reduces computational complexity. Many
applications, however, involve nanobeams of finite thickness with
exposed side surfaces. To examine how exposed
surfaces change the optimum, we use Nano-TO to design finite-thickness
nanocantilevers.

The design domain is 200.475$\times$60.75$\times$615.60~\AA\ and
contains 451{,}440 atoms: 445{,}500 active atoms and 5{,}940 passive
atoms at the clamped boundary. Compared to the thickness-periodic case,
we remove PBCs in the thickness direction and triple the thickness; all
other settings (two-phase update schedule, filtering) are identical. We
perform 16 independent trials. Figure~S12 shows the initial design
(100\%) and the best designs at mass ratios of 75.76\%, 67.68\%, and
59.60\%. The optimized designs feature nearly closed-wall motifs.
Figure~S13 plots normalized bending stiffness over all trials
(colored circles), relative to the initial design. As baselines, we
include height-scaled reference beams with stiffnesses obtained from
atomistic simulations (blue dots) and estimated using the
Euler--Bernoulli theory (red curve). Across mass ratios, the optimized
designs consistently exceed the references at equal mass.

Figure~S14a shows the optimized design with a mass ratio of 59.60\%,
color-coded by coordination number. For reference, we triple the
thickness of the thickness-periodic design from the main text and
remove PBCs. Figure~S14b shows that the exposed surfaces are
predominantly \{100\} with a coordination number of 8.
Figures~S14c--f map the normal ($\eps_{zz}$) and shear
($\eps_{xz}$) strains. Both nanocantilevers are in tension at the top
and compression at the bottom, inducing shear across the section. In
the finite-thickness design (Figures~S14c,e), shear spreads through a
continuous wall, reducing localized strain concentrations. In the
scaled thickness-periodic design (Figures~S14d,f), cross-braces convert
shear into axial forces along their lengths, concentrating shear
near brace nodes and creating more localized strain ``hotspots,''
leading to lower bending stiffness than its finite-thickness
counterpart.

\begin{figure}[H]
  \centering
  \includegraphics[width=0.92\textwidth]{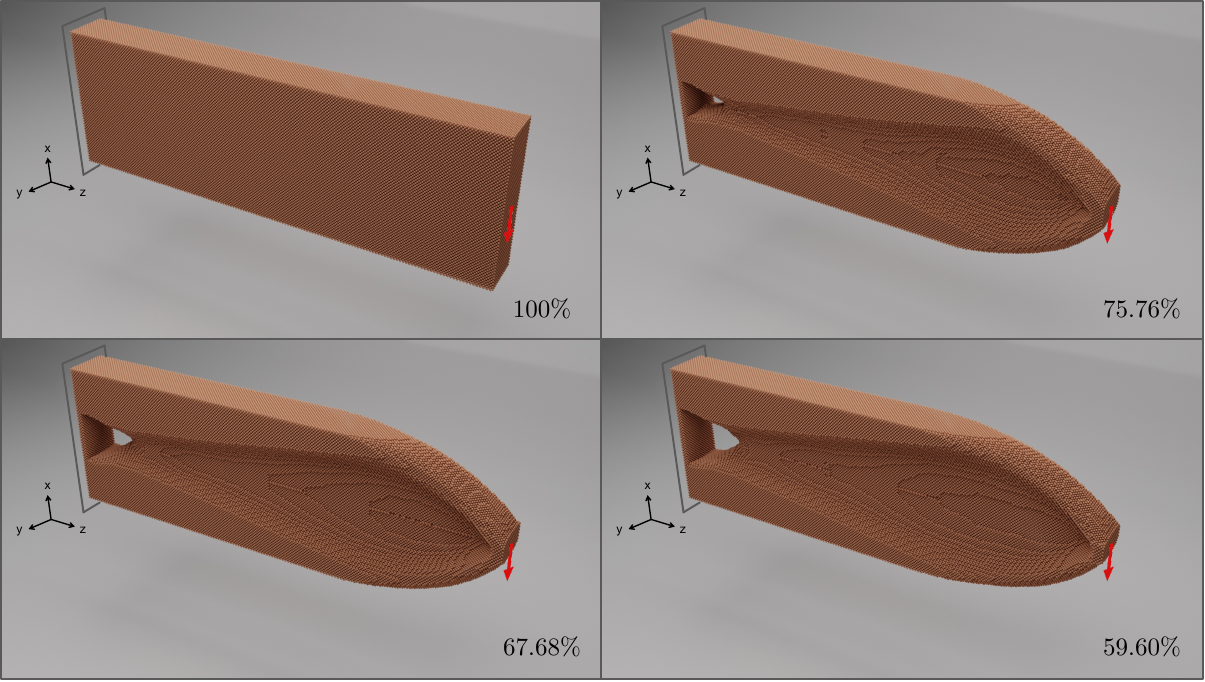}
  \caption{\textbf{Nano-TO design of finite-thickness nanocantilevers.}
    Initial design (100\%) and optimized designs at different mass
    ratios (75.76\%, 67.68\%, and 59.60\%). The gray block represents
    the clamped support; the red arrow marks the applied vertical
    displacement at the free end.}
  \label{fig:S12}
\end{figure}

\begin{figure}[H]
  \centering
  \includegraphics[width=0.62\textwidth]{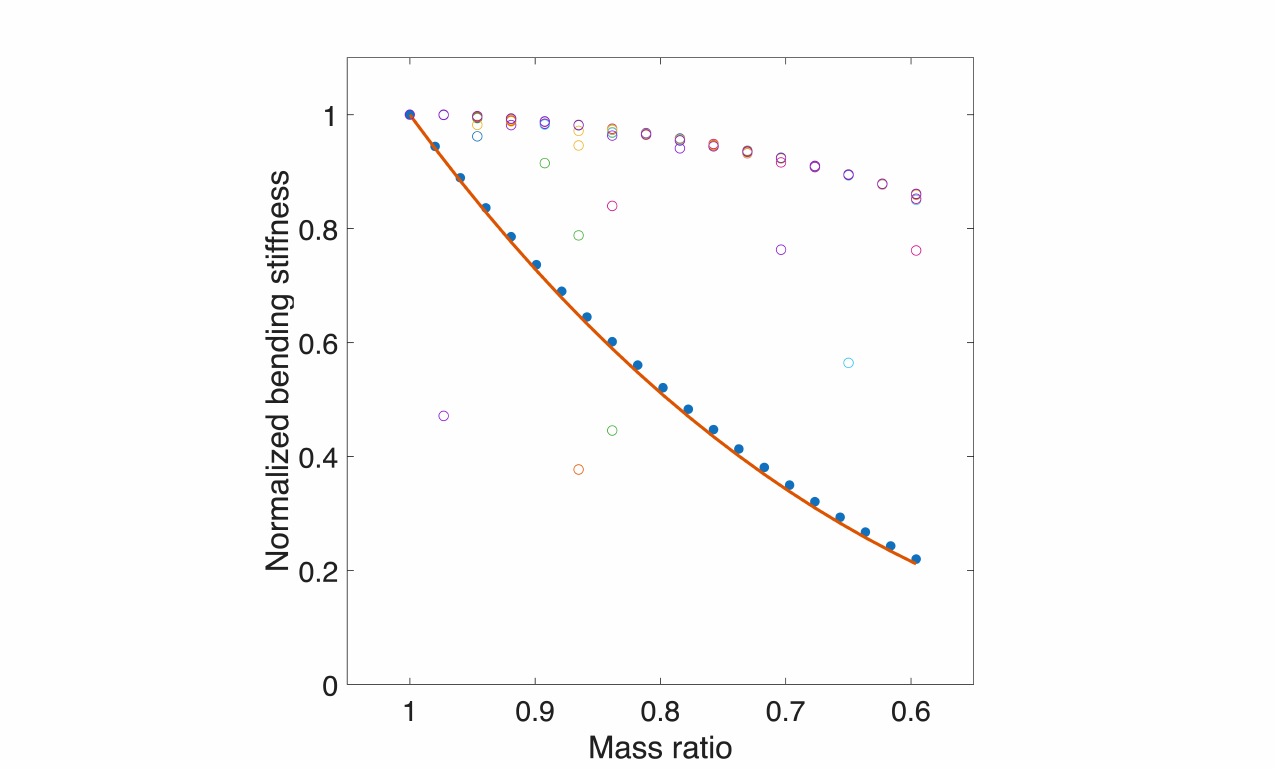}
  \caption{\textbf{Normalized bending stiffness of finite-thickness
    nanocantilevers versus mass ratio.} Colored circles: results from 16
    independent trials at each mass ratio. Blue dots: height-scaled
    reference designs. Red curve: Euler--Bernoulli estimate.}
  \label{fig:S13}
\end{figure}

\begin{figure}[H]
  \centering
  \includegraphics[width=0.92\textwidth]{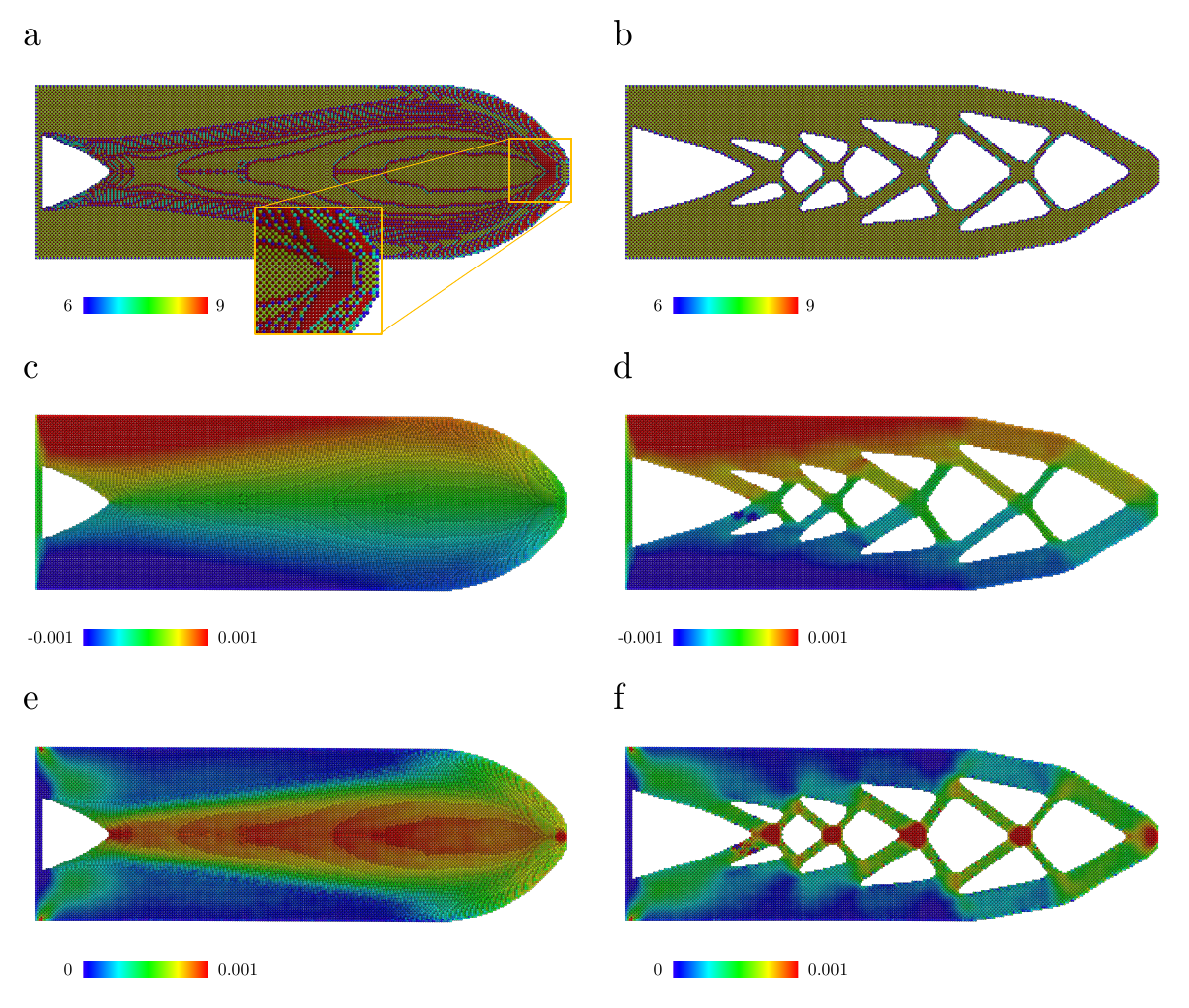}
  \caption{\textbf{Finite-thickness versus thickness-periodic
    nanocantilevers.}
    \textbf{a}, Optimized finite-thickness design at a mass ratio of
    59.60\%, colored by coordination number (6--9). \textbf{b},
    Optimized thickness-periodic design with exposed \{100\} surfaces
    (coordination 8). \textbf{c} and \textbf{d}, Normal strain
    ($\eps_{zz}$) map when a vertical displacement is applied at the
    free end. \textbf{e} and \textbf{f}, Shear ($\eps_{xz}$) strain
    map. Color bars indicate the strain ranges.}
  \label{fig:S14}
\end{figure}

\subsection*{A.12\quad Bulk-equivalent bending stiffness reference}

For finite-thickness nanocantilevers with traction-free side surfaces,
the measured bending stiffness depends on thickness because a
non-negligible fraction of atoms resides near the side surfaces at small
thickness. To quantify the stiffness penalty associated with exposed
side surfaces, we compare the thickness-normalized (effective) bending
stiffness of a thin nanobeam to a bulk-equivalent reference representing
the interior response in the large-thickness limit.

Using a single very thick beam (e.g., 100$\times$ the original
thickness) as a bulk reference can still underestimate surface effects,
as the thickness-normalized stiffness may remain several percent below
its thick-limit value. We therefore estimate the bulk-equivalent
reference from the linear scaling of total stiffness with thickness in
the thick regime (50$\times$ to 100$\times$), which effectively isolates
the interior contribution.

Let the beam thickness be $t = nt_0$, where $t_0 = 20.25$~\AA\ is the
base thickness (1$\times$) and $n$ is the thickness multiplier. All
beams in this note share the same in-plane topology (the
thickness-periodic Nano-TO design with a mass ratio of 59.60\%) and the
same loading and boundary conditions, except for thickness. For each
thickness multiplier $n$, we impose the same bending deformation
used in the main text and relax the atomic positions. We define an
energy-based bending stiffness proxy $K(n)$ (units: eV) as the
minimized elastic energy increment under this fixed-displacement
loading. Because the imposed displacement amplitude is identical for
all designs compared in this note, $K(n)$ is proportional to the
effective bending stiffness and can be used to compare designs and
thicknesses on a consistent basis.

To compare different thicknesses, we define the effective bending
stiffness $k(n) = K(n)/t = K(n)/(nt_0)$. Since $t_0$ is constant, we
report $k(n) \propto K(n)/n$. In a homogeneous continuum beam without
surface effects, $k(n)$ is independent of $n$. In atomistics with
traction-free side surfaces, $k(n)$ varies with $n$ as the
surface-atom fraction decreases.

Once the two side surfaces are sufficiently separated such that their
local response is thickness-independent, the total stiffness can be
decomposed into an interior term proportional to thickness plus a
thickness-independent surface correction:

\begin{equation}
  K(n)
  \approx \kbulk \cdot n t_0 - C
  \tag{S6a}
\end{equation}

where $\kbulk$ is the bulk-equivalent stiffness per thickness, and
$C > 0$ is a constant capturing the net reduction caused by the two side
surfaces. Dividing by the thickness multiplier yields:

\begin{equation}
  \frac{K(n)}{n}
  \approx \kbulk \cdot t_0 - \frac{C}{n}
  = \kbulkstar - \frac{C}{n}
  \tag{S6b}
\end{equation}

where $\kbulkstar$ is the bulk-equivalent stiffness per baseline
thickness. We compute $K(n)$ for thick beams with traction-free side
surfaces at $n \in \{50, 60, 70, 80, 90, 100\}$ and fit $K(n)$ to a
linear function of $n$: $K(n) = \kbulkstar \cdot n - C$. Using
least-squares regression, we obtain $\kbulkstar = 0.320877$ and $C =
1.090833$. This linear model describes the 50$\times$ to 100$\times$
data extremely well, indicating that $n \geq 50$ is within the thick
regime for estimating $\kbulkstar$. Table~S2 lists the values used in
the fit, the predicted values, and the relative error.

We define the side-surface penalty at thickness multiplier $n$ as the
fractional reduction in stiffness relative to the bulk-equivalent
reference:

\begin{equation}
  \mathrm{Penalty}(n)
  = 1 - \frac{K(n)}{\kbulkstar \times n}
  \tag{S7}
\end{equation}

For the 3$\times$ beam, $K(3) = 0.8379$, therefore
$\mathrm{Penalty}(3) = 0.12964 \approx 13.0\%$. This 13\% value
quantifies the reduction in effective bending stiffness attributable to
exposed side surfaces at 3$\times$ thickness for this fixed topology.

For the thickness-periodic beam, the effective stiffness is
$k_\mathrm{PBC} = 0.321059$. This agrees with the bulk-equivalent
reference within 0.06\%, indicating that the periodic cross-section assumption, including the
out-of-plane kinematic constraint and the absence of side surfaces,
introduces negligible bias in the bending stiffness for this topology
and loading.

\begin{table}[H]
  \centering
  \caption{\textbf{Energy-based stiffness proxy for thick beams used to
    estimate the bulk-equivalent stiffness.} $K_{\mathrm{actual}}$
    (eV) is obtained from atomistic simulations, and $K_{\mathrm{pred}}$
    (eV) is given by the least-squares linear fit over thickness
    multipliers $n = 50$ to 100. Error is
    $(K_{\mathrm{pred}} - K_{\mathrm{actual}})/K_{\mathrm{actual}} \times 100\%$.}
  \label{tab:S2}
  \begin{tabular}{lcccccc}
    \toprule
    $n$                     & 50       & 60       & 70       & 80       & 90       & 100      \\
    \midrule
    $K_{\mathrm{actual}}$   & 14.9545  & 18.1575  & 21.3748  & 24.5786  & 27.7866  & 30.9977  \\
    $K_{\mathrm{pred}}$     & 14.9530  & 18.1618  & 21.3706  & 24.5793  & 27.7881  & 30.9969  \\
    Error (\%)              & $-0.0100$ & 0.0237   & $-0.0197$ & 0.0028  & 0.0054   & $-0.0026$ \\
    \bottomrule
  \end{tabular}
\end{table}

\subsection*{A.13\quad Design of finite-thickness nanocantilevers at reduced scale}

To study size dependence, we re-optimize finite-thickness
nanocantilevers with all dimensions reduced to approximately 40\% of
those in Section~A.11. The design domain is
79.975$\times$24.30$\times$251.10~\AA\ with 29{,}016 atoms: 28{,}080
active atoms and 936 passive atoms at the clamped boundary. To keep the
relative minimum feature size comparable, the filter radius is reduced
from 10.325~\AA\ to 4.05~\AA. All other settings, including the
two-phase update schedule, are unchanged. We perform 64 independent
trials.

Figure~S15 shows the initial design (100\%) and the best designs at mass
ratios of 75.78\%, 67.95\%, and 60.11\%. At a higher mass ratio (e.g.,
75.78\%), the optimized design remains a nearly closed wall. As the mass
ratio decreases, the wall becomes only a few atomic layers thick, making
it an unstable carrier of transverse shear. Therefore, the layout
transforms into a truss-like configuration at a lower mass ratio (e.g.,
60.11\%). Figure~S16 plots normalized bending stiffness over all trials
(colored circles), relative to the initial design. As baselines, we
include height-scaled reference beams (also with all dimensions
reduced), with normalized bending stiffness obtained from atomistic
simulations (blue dots) and estimated using the Euler--Bernoulli theory
(red curve). Across mass ratios, the optimized designs consistently
exceed the references at equal mass.

\begin{figure}[H]
  \centering
  \includegraphics[width=0.92\textwidth]{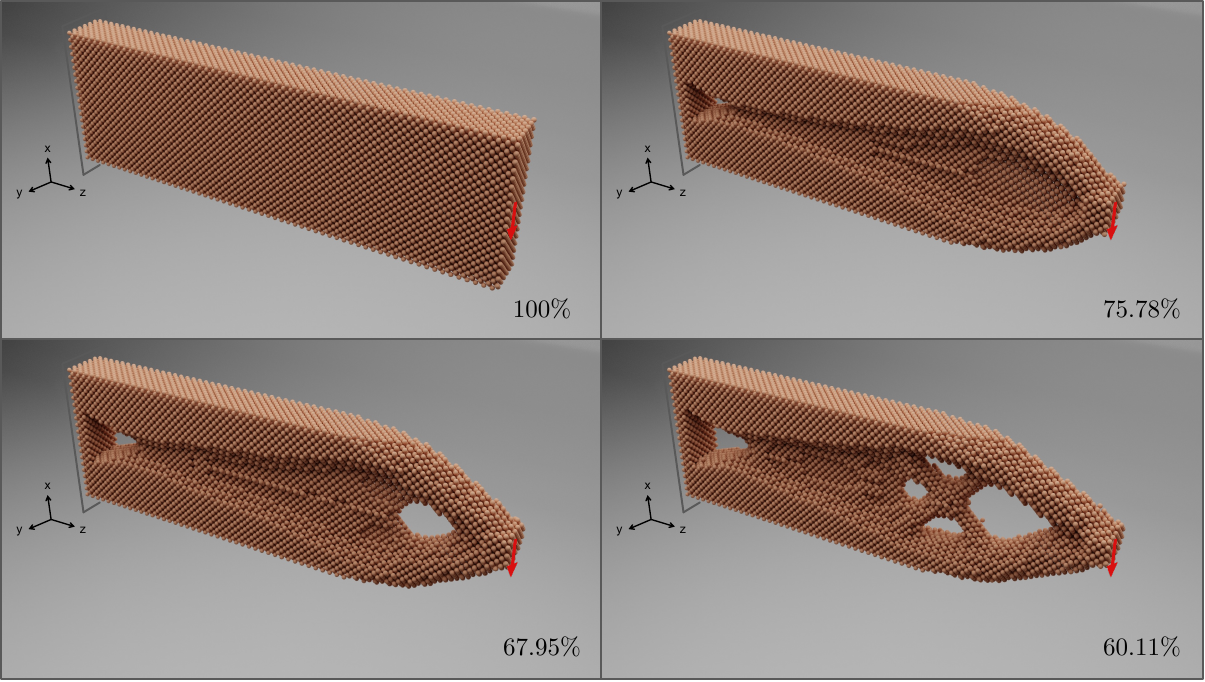}
  \caption{\textbf{Nano-TO design of finite-thickness nanocantilevers at
    reduced scale.} Initial beam (100\%) and optimized designs at
    different mass ratios (75.78\%, 67.95\%, and 60.11\%). The gray
    block represents the clamped support; the red arrow marks the
    applied vertical displacement at the free end.}
  \label{fig:S15}
\end{figure}

\begin{figure}[H]
  \centering
  \includegraphics[width=0.62\textwidth]{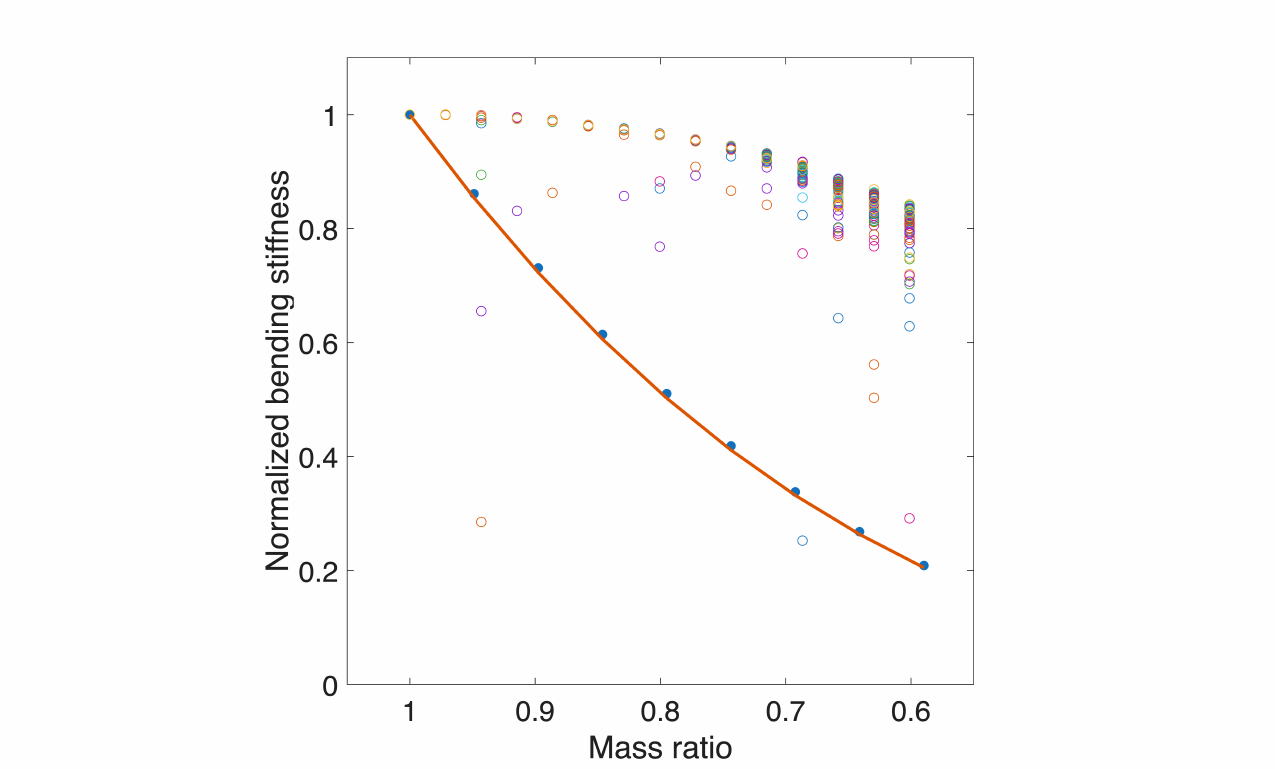}
  \caption{\textbf{Normalized bending stiffness of finite-thickness
    nanocantilevers at reduced scale versus mass ratio.} Colored
    circles: results from 64 independent trials at each mass ratio. Blue
    dots: height-scaled reference designs. Red curve: Euler--Bernoulli
    estimate.}
  \label{fig:S16}
\end{figure}

\subsection*{A.14\quad Nanopillars designed by FEM-TO versus Nano-TO}

Classical topology optimization based on the finite element method
(FEM-TO) treats solids as homogeneous media and ignores surface
elasticity and facet specificity. To quantify the advantages of
explicitly modeling atoms, we design the same nanopillar using FEM-TO
and evaluate its vertical stiffness as an atomistic structure under
identical loading and boundary conditions. The FEM-TO method is based on
the 169-line MATLAB code from Liu and Tovar~\cite{Liu2014}, available
at \url{http://top3dapp.com}.

For Nano-TO, the nanopillar design domain is generated from
$40\times40\times102$ conventional FCC cubic unit cells, with the cube
edges along $[100]/[010]/[001]$. The first two layers of unit cells in
the $[001]$ direction contain only passive atoms and are not involved
in the design process. These atoms represent the anchoring base made of
the same material. Without them, the bottom of the nanopillar would be
exposed surfaces with different elastic properties. As shown in
Figure~S17, each clamped support is a square with a length equal to
20\% of the base length, thus representing 4\% of the base area, and
can be mapped to a finite element mesh with $8\times8$ elements.

Since there are no surface effects in FEM, the bottom two layers are not
included in FEM-TO, giving a $40\times40\times100$ element mesh. To
better compare FEM-TO and Nano-TO, we set FEM-TO parameters to match
the physical properties in Nano-TO. The Poisson's ratio is set to
0.351, calculated using the same EAM potential as that used in
Nano-TO~\cite{Mishin1999}. The filter radius is set to 2.549,
matching the ratio of the Nano-TO filter radius to the unit cell size
(10.325/4.05). The target mass ratio is set to 20.25\%, the same as
that in Nano-TO. The penalty is set to 3. The solid modulus is set to
1.0, and the void modulus is set to $10^{-9}$. The convergence
tolerance (design change) is set to $10^{-3}$.

The optimized design from FEM-TO is converted to a binary layout, as
shown in Figure~S18. This layout is voxelized onto the FCC lattice,
where each ``solid'' voxel becomes a unit cell with four atoms. To
ensure a fair comparison with the Nano-TO design in the main text, the
same base consisting of two layers of unit cells is added to the model
(Figure~S18). The FEM-TO design has a normalized vertical stiffness of
3.44. In comparison, the Nano-TO design reaches a normalized vertical
stiffness of 3.65, which is 6.10\% higher than the FEM-TO design.

Lastly, we begin with the FEM-TO design and subsequently optimize it
using Nano-TO. Figure~S19 shows the evolution of the nanopillar over
10, 100, and 1{,}000 iterations. From the initial design, composed of
cubic unit cells with the cube edges along $[100]/[010]/[001]$, it
evolves into a smoother design featuring a variety of surfaces. The
Nano-TO re-optimized design has a normalized vertical stiffness of
3.70, 7.40\% higher than that of the FEM-TO design. Interestingly,
this new optimized design is also slightly stiffer than the Nano-TO
design reported in the main text (3.70 vs.\ 3.65).

\begin{figure}[H]
  \centering
  \includegraphics[width=0.77\textwidth]{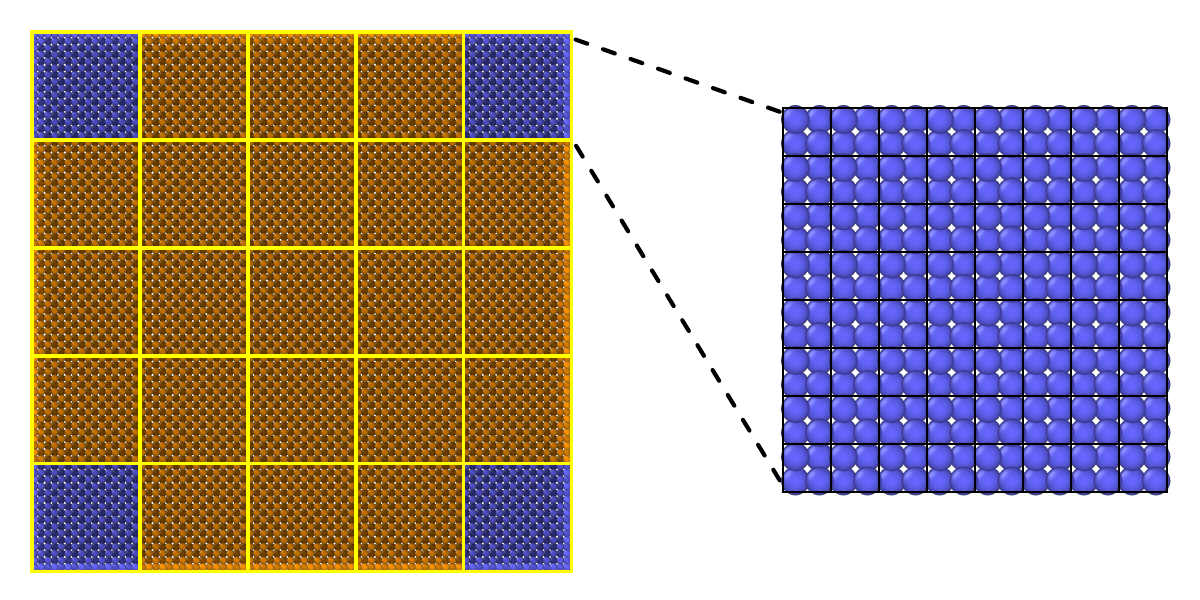}
  \caption{\textbf{Supports of nanopillars in atomistic and finite
    element models.} Each clamped support (blue) is a square with a
    length equal to 20\% of the base length, representing 4\% of the
    base area (left). This boundary condition can be mapped to a finite
    element mesh with $8\times8$ elements (right).}
  \label{fig:S17}
\end{figure}

\begin{figure}[H]
  \centering
  \includegraphics[width=0.46\textwidth]{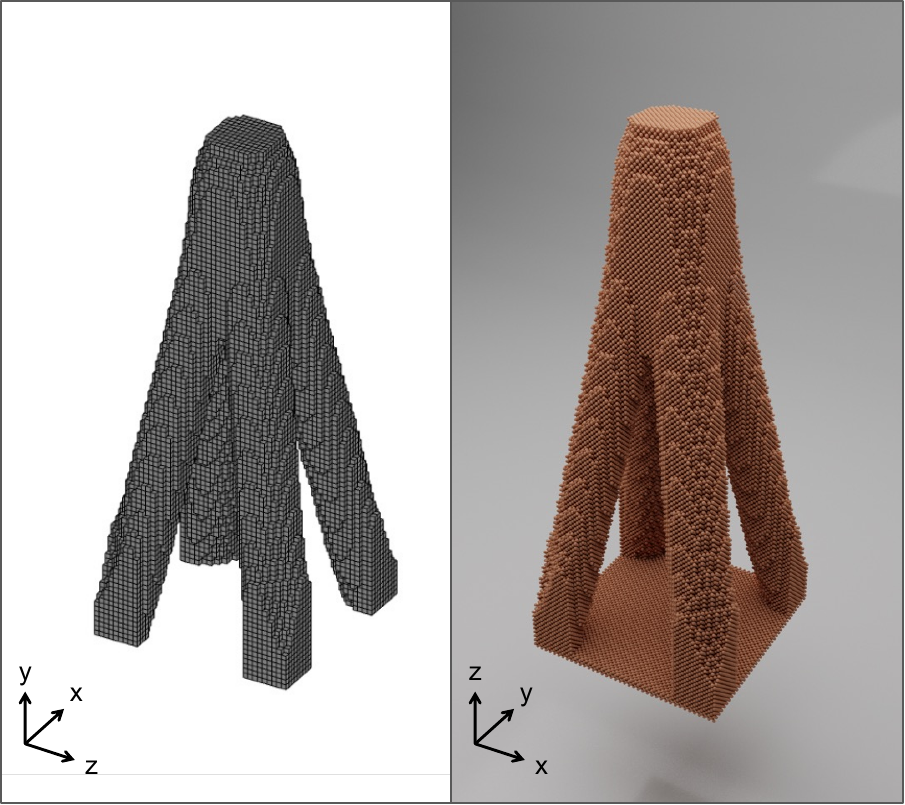}
  \caption{\textbf{Binary FEM-TO design and its corresponding atomistic
    model.} The optimized design from FEM-TO is converted to a binary
    layout (left). This layout is voxelized onto the FCC lattice, where
    each ``solid'' voxel becomes a unit cell with four atoms (right).
    The orientations of the coordinate systems are adjusted during
    mapping.}
  \label{fig:S18}
\end{figure}

\begin{figure}[H]
  \centering
  \includegraphics[width=0.92\textwidth]{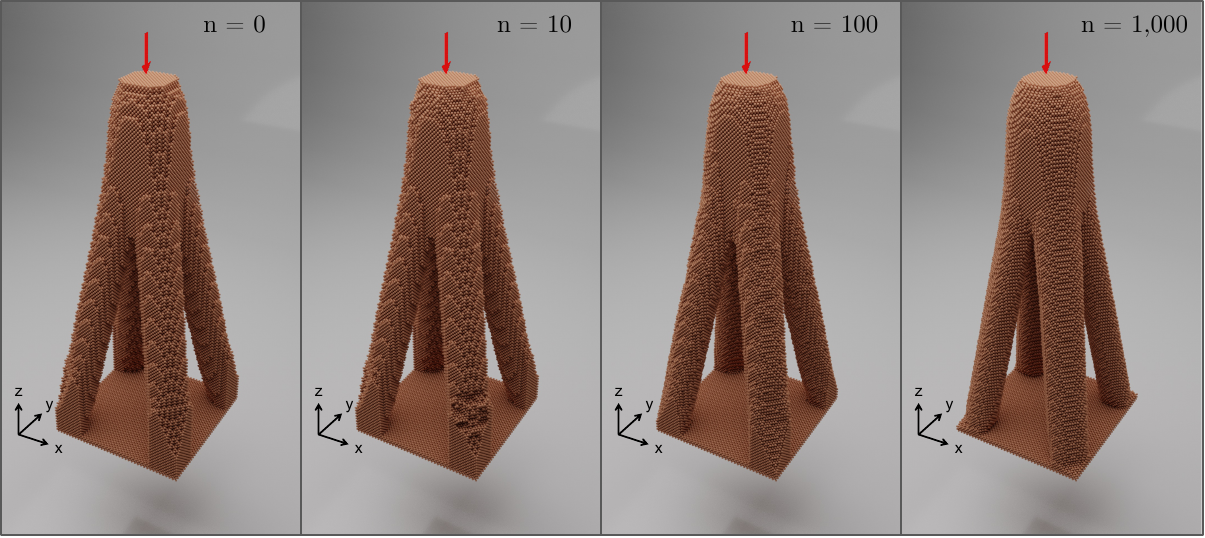}
  \caption{\textbf{Nano-TO design of nanopillars using an initial design
    from FEM-TO.} Initial pillar from FEM-TO and optimized designs at
    different iterations (10, 100, and 1{,}000). The red arrow marks the
    applied vertical displacement at the center of the top surface.}
  \label{fig:S19}
\end{figure}

\ifdefined\SIincluded\else
\subsection*{A.15\quad Supplementary references}

\begingroup
\renewcommand{\section}[2]{}%
\bibliographystyle{unsrtnat}

\endgroup

\end{document}
\fi

\end{document}